\documentclass[12pt]{article}
\pdfoutput=1
\usepackage{putex,hyperref}
\usepackage{mathtools}
\usepackage{graphicx}
\usepackage{caption}
\usepackage{subcaption}
\usepackage{epstopdf}
\usepackage{enumerate}
\usepackage{cite}
\usepackage{youngtab}
\usepackage{tensor}
\usepackage{slashed}
\usepackage[aligntableaux=center]{ytableau}

\numberwithin{equation}{section}

\newcommand {\be} {\begin {equation}}
\newcommand {\ee} {\end {equation}}

\newcommand {\bes} {\begin {equation*}}
\newcommand {\ees} {\end {equation*}}

\newcommand{\es}[2] {\begin{equation} \label{#1} \begin{split} #2 \end{split} \end{equation}}

\newcommand{\Z}{\mathbb{Z}}

\newcommand{\R}{\mathbb{R}}
\newcommand{\C}{\mathbb{C}}

\def\Tr{\mop{Tr}}

\newcommand{\beq}{\begin{equation}}
\newcommand{\eeq}{\end{equation}}

\newcommand{\del}{\partial}

\def\SSP#1{{}}
\def\RY#1{}
\def\MD#1{}

\def\<{\langle}
\def\>{\rangle}

\newcommand{\cA}{\ensuremath{\mathcal{A}}}

\newcommand{\cD}{\ensuremath{\mathcal{D}}}

\newcommand{\cH}{\ensuremath{\mathcal{H}}}

\newcommand{\cM}{\ensuremath{\mathcal{M}}}
\newcommand{\cN}{\ensuremath{\mathcal{N}}}
\newcommand{\cO}{\ensuremath{\mathcal{O}}}
\newcommand{\cP}{\ensuremath{\mathcal{P}}}
\newcommand{\cQ}{\ensuremath{\mathcal{Q}}}
\newcommand{\cR}{\ensuremath{\mathcal{R}}}

\newcommand{\cV}{\ensuremath{\mathcal{V}}}
\newcommand{\cW}{\ensuremath{\mathcal{W}}}

\newcommand{\dd}{\mathrm{d}}

\newcommand{\ed}{\,.}
\newcommand{\ec}{\,,}

\newcommand{\tpsi}{\widetilde{\psi}}
\newcommand{\tq}{\widetilde{q}}
\newcommand{\tQ}{\widetilde{Q}}

\begin{document}
	\institution{Fudan}{Center for Mathematics and Interdisciplinary Sciences,\cr Fudan University, Shanghai, 200433, China.}
	\institution{SIMIS}{Shanghai Institute for Mathematics and Interdisciplinary Sciences (SIMIS),\cr Shanghai, 200433, China.}
	\institution{YITP}{C.N. Yang Institute for Theoretical Physics,\cr Stony Brook University, Stony Brook, NY 11794, USA.}

	\title{Coulomb Branches of Noncotangent Type:\\ a Physics Perspective}
	\authors{Mykola Dedushenko\worksat{\Fudan,\SIMIS} and Daniel Resnick\worksat{\YITP}}
	
	\abstract{We study the Coulomb-branch sector of 3D $\mathcal{N}=4$ gauge theories with half-hypermultiplets in general pseudoreal representations $\mathbf{R}$ (``noncotangent'' theories).  This yields (short) quantization of the Coulomb branch and correlators of the Coulomb branch operators captured by the 1d topological sector. This is done by extending the hemisphere partition function approach to noncotangent matter.  In this setting one must first cancel the parity anomaly, and overcome an obstacle that $(2,2)$ boundary conditions for half-hypers are generically incompatible with gauge symmetry. Using the Dirichlet boundary conditions for the gauge fields and a careful treatment of half-hypermultiplet boundary data, we describe the resulting shift/difference operators implementing monopole insertions (including bubbling effects) on $HS^3$, and use the $HS^3$ partition function as a natural module on which the Coulomb-branch operator algebra $\mathcal{A}_C$ is represented.  As applications we derive generators and relations of $\mathcal{A}_C$ for $SU(2)$ theories with general matter (including half-integer spin representations), analyze theories with Coulomb branch $y^2=z(x^2-1)$, compute the Coulomb branch of an $A_n$ quiver with spin-$\frac32$ half-hypers, and check consistency of a general monopole-antimonopole two-point function.}
	
	\date{}
	
	\maketitle
	
	\tableofcontents

	\setlength{\unitlength}{1mm}
	
	\newpage

\section{Introduction}\label{sec:Intro}
Three-dimensional theories with $\cN=4$ supersymmetry host a large body of exactly solvable structure.  This includes moduli spaces of vacua \cite{Hanany:1996ie} (in particular, Higgs and Coulomb branches), various dualities, including 3D mirror symmetry \cite{Intriligator:1996ex,deBoer:1996ck,Kapustin:1999ha} and symplectic duality \cite{Braden2012QuantizationsOC,Bullimore:2016nji}, and supersymmetric partition functions and correlation functions \cite{Kapustin:2009kz,Kapustin:2010mh,Dedushenko:2016jxl,Dedushenko:2017avn,Dedushenko:2018icp}.  The Coulomb branch is especially subtle: unlike the (purely classical) Higgs branch \cite{Hitchin:1986ea}, it is shaped by strong quantum effects \cite{Seiberg:1996bs,Seiberg:1996nz,Kapustin:1999ha,Hanany:1996ie,Gaiotto:2008ak}, and in a Lagrangian description it is naturally probed by disorder defects---the monopole operators \cite{Borokhov:2002ib,Borokhov:2002cg,Borokhov:2003yu,Bashkirov:2010kz}.  Defining and controlling QFT in the presence of disorder operators is intrinsically harder than for ordinary local (order) operators, and often involves strong-coupling phenomena.

For that reason, a proper description of Coulomb branches appeared relatively recently. Motivated in part by earlier physical insights such as the monopole formula \cite{Cremonesi:2013lqa}, it was pioneered in the mathematical literature \cite{Nakajima:2015txa,Braverman:2016wma,Braverman:2016pwk}.  A more physics-transparent viewpoint was later developed via abelianization \cite{Bullimore:2015lsa} and via supersymmetric localization \cite{Dedushenko:2017avn,Dedushenko:2018icp} (which ultimately justifies abelianization).  These approaches, however, primarily addressed 3D $\cN=4$ Lagrangian theories of \emph{cotangent} type, namely theories whose hypermultiplet matter is a cotangent bundle $T^*\cR$ (equivalently, half-hypermultiplets in $\cR\oplus\overline{\cR}$, or full hypermultiplets in $\cR$).  In contrast, in \emph{noncotangent} theories the matter consists of half-hypermultiplets valued in a general pseudoreal representation $\mathbf{R}$ that is not of the form $T^*\cR$.  Their Coulomb branches remained less accessible, until a recent mathematical work \cite{Braverman:2022zei}.  The goal of this paper is to generalize the localization-based techniques of \cite{Dedushenko:2017avn,Dedushenko:2018icp} to the noncotangent setting.

Once we allow a general pseudoreal $\mathbf{R}$, two new issues arise that are absent when $\mathbf{R}=\cR\oplus\overline{\cR}$.  The first is the $\Z_2$-anomaly, i.e.\ the 3D parity anomaly \cite{Niemi:1983rq,Redlich:1983kn,Redlich:1983dv}.  For full hypermultiplets this anomaly cancels automatically, but for general $\mathbf{R}$ it need not.  In non-supersymmetric 3D gauge theory one can cancel the parity anomaly by adding a Chern--Simons (CS) counterterm, but in an $\cN=4$ theory such a term typically breaks $\cN=4$ down to $\cN=3$ \cite{Kao:1992ig,Gaiotto:2007qi}.  We therefore restrict attention to non-anomalous theories, in which the matter in $\mathbf{R}$ cannot generate half-integral CS terms at low energies (see Section \ref{sec:anomaly}).  The anomaly cancellation condition can also be phrased (see \cite[Appendix B]{Braverman:2022zei}) as the vanishing of the map $\pi_4 G \to \pi_4 {\rm Sp}(\mathbf{R})$, where $G\to {\rm Sp}(\mathbf{R})$ is the given representation written so as to exhibit the quaternionic structure required by eight supercharges.  This is the familiar condition for canceling Witten's $SU(2)$ anomaly in 4D \cite{Witten:1982fp}: a theory suffering from the 4D $SU(2)$ anomaly formally reduces to a 3D theory with a parity anomaly.  Because of this relation, the parity anomaly of half-hypermultiplets is sometimes informally referred to as Witten's anomaly, even though the former is a 3D effect and the latter is intrinsically 4D and makes the theory inconsistent.  (Unlike Witten's anomaly, the parity anomaly can be canceled by counterterms; in our $\cN=4$ setting such counterterms would typically reduce supersymmetry.)

The second subtlety becomes manifest when we study $(2,2)$ boundary conditions \cite{Chung:2016pgt,Bullimore:2016nji,Dimofte:2011ju}.  For (half-)hypers such boundary conditions require choosing a splitting $\mathbf{R}\cong\mathbb{L}\oplus \mathbb{L}^\perp$, where $\mathbb{L}$ and $\mathbb{L}^\perp$ are transversal Lagrangian subspaces.  For full hypers, $\mathbf{R}=\cR\oplus\overline\cR$ provides a natural splitting compatible with gauge symmetry.  By contrast, the defining feature of a genuine half-hypermultiplet is precisely that no such $G$-invariant Lagrangian splitting exists.  Thus half-hypermultiplets do not admit $G$-invariant $(2,2)$ boundary conditions.  This is a problem for Neumann boundary conditions on the gauge fields, since those must preserve gauge symmetry along the boundary.\footnote{One can contemplate resolving this issue by imposing boundary conditions in a gauge-fixed theory, but we then expect the problematic boundary conditions to violate BRST symmetry or be otherwise inconsistent.}  Dirichlet boundary conditions on the gauge fields, however, are compatible with explicitly breaking $G$ at the boundary (where it becomes a global symmetry).  Since the methods of \cite{Dedushenko:2017avn,Dedushenko:2018icp} rely on $(2,2)$ boundary conditions but use only Dirichlet boundary conditions for the gauge fields, we can proceed, though we will need some care in deriving the analog of the gluing formula relating the $S^3$ and $HS^3$ partition functions.

\subsubsection*{Technical overview}
\begin{figure}[h]
	\centering
	\includegraphics[width=0.4\textwidth]{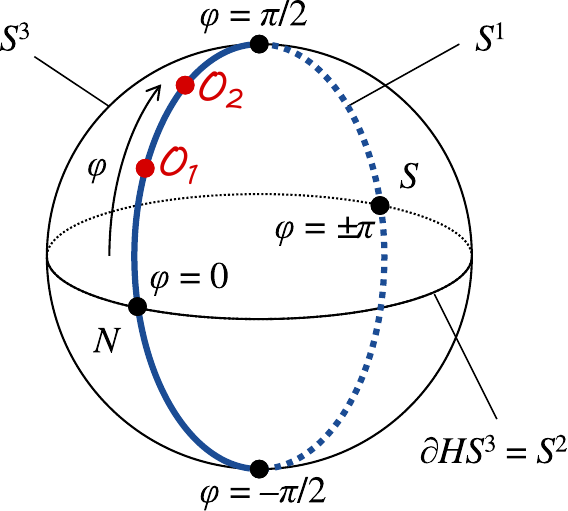}
	\caption{Along the fixed circle $S^1_\varphi \subset S^3$, one can insert the twisted-translated Higgs or Coulomb branch operators to compute the protected correlation functions.}\label{fig:S3geom}
\end{figure}

Let us briefly review the approach to Coulomb branches developed in \cite{Dedushenko:2017avn,Dedushenko:2018icp} and explain what must be modified for noncotangent theories. In \emph{loc. cit.}, the Coulomb sector $\cA_C$ is accessed through supersymmetric localization on the hemisphere $HS^3$ with $(2,2)$ boundary conditions.\footnote{Hemisphere partition functions have been long known to be useful in 2D physics \cite{Hori:2013ika,Knapp:2016rec,Hori:2019vkm}, but now we also learn about their usefulness in 3D, like in the current paper and earlier references.} The localization is done with respect to the supercharge $\cQ_C$ (compatible with the boundary conditions) that squares to a $U(1)_J$ rotation of $HS^3$ plus a $U(1)_H\subset SU(2)_H$ R-symmetry transformation. The fixed locus of $U(1)_J$ on the full sphere is a circle $S^1\subset S^3$, and on the hemisphere, it is an interval (half-circle) $HS^1\subset HS^3$ (see Figure \ref{fig:S3geom}). The cohomology of $\cQ_C$ contains local Coulomb branch operators inserted along this $S^1$ or $HS^1$. At one point $0\in S^1$, they form the Coulomb branch chiral ring (in one complex structure), however, as we move along the $S^1$, they undergo a nontrivial $SU(2)_C$ rotation. Such twisted Coulomb branch operators form a one-dimensional topological sector, first defined in a flat space SCFT in \cite{Chester:2014mea,Beem:2016cbd}. The algebra $\cA_C$ of these local operators gives a short quantization \cite{Etingof:2019guc,Etingof:2020fls,Klyuev:2025cau,Klyuev:2025fxl} of the Coulomb branch, with $\frac{1}{r}$, the inverse radius of $S^3$, being the quantization parameter. The commutative limit of $\cA_C$, which means the flat space limit $r\to\infty$, recovers the coordinate ring $\C[X_C]$ of the Coulomb branch.

The localization allows to compute hemisphere partition function decorated by the insertions of twisted Coulomb branch operators along the $HS^1\subset HS^3$. One can further recover the SCFT correlators by gluing in the second hemisphere to get the full $S^3$ with insertions as in Figure \ref{fig:gluing-sketch}, which is well-defined in a ``good'' or ``ugly'' theory \cite{Gaiotto:2008ak}.  However, our hemisphere construction provides the algebra $\cA_C$ and its representation on the hemisphere independently of the ``good/ugly/bad'' status of the theory, which makes it especially powerful.

The $(2,2)$ boundary conditions on the vector multiplet on $HS^3$ are parameterized in terms of $(\sigma, B)$, where $\sigma\in\mathfrak{t}$ is a Cartan-valued real boundary mass (boundary value of the vector multiplet scalar), and $B\in\Lambda^\vee\subset\mathfrak{t}$ is a coweight representing magnetic flux through the boundary (recall that non-abelian magnetic fluxes live in the coweight lattice $\Lambda^\vee$, which is the weight lattice of Langlands or GNO dual group \cite{Goddard:1976qe}), see eqn. \eqref{eq:VM-bc}. For the half-hypermultiplets, the boundary conditions are labeled by the Lagrangian splitting $\mathbf{R}=\mathbb{L}\oplus \mathbb{L}^\perp$, see eqn. \eqref{eq:HM-bc}.  The resulting hemisphere partition function is called a hemisphere wavefunction, and it depends on $(\sigma, B)$ and $\mathbb{L}$:
\begin{equation}
Z_{HS^3}(\sigma,B; \mathbb{L}) \in Fun(\mathfrak{t}\times \Lambda^\vee, \C),
\end{equation}
where we intentionally leave the precise function space implicit. In fact, it is smooth in real variable $\sigma$ (but has poles for complex $\sigma$). The empty hemisphere partition function is:
\begin{equation}\label{eq:emptyZHS3}
Z_{HS^3}(\sigma, B; \mathbb{L})=\delta_{B,0}\frac{\prod_{w\in\mathbb{L}} \frac1{\sqrt{2\pi}} \Gamma\left(\frac12 - i w\cdot\sigma\right)}{\prod_{\alpha\in\Delta} \frac1{\sqrt{2\pi}}\Gamma(1 - i\alpha\cdot\sigma)}.
\end{equation}
Here $\Delta$ is the set of all nonzero roots of $G$, and we also abuse the notations by denoting the set of weights in $\mathbb{L}\subset \mathbf{R}$ by the same symbol $\mathbb{L}$.

In this formalism, monopole operator insertions act on $Z_{HS^3}(\sigma,B;\mathbb{L})$ by shift/difference operators in $(\sigma,B)$, and one obtains an explicit action of $\cA_C$ on the space of hemisphere wavefunctions. In this way we generate an $\cA_C$-module as
\begin{equation}
\mathbb{M}=\cA_C\, Z_{HS^3}(\sigma,B;\mathbb{L}).
\end{equation}
We define the algebra $\cA_C$ through its action on $\mathbb{M}$, so this representation is faithful by definition, i.e.\  the relations satisfied in the module $\mathbb{M}$ are precisely the relations in $\cA_C$. This is reasonable because correlators in the protected 1D Coulomb-branch sector are recovered from this representation, and relations ``under the correlators'' is what defines the operator algebra in QFT.

We will find two types of generators in $\cA_C$: scalars that act on $Z_{HS^3}$ diagonally, simply multiplying it by $\Phi=\frac{1}{r} \left(\sigma + \frac{i}{2}B\right)$, and BPS monopoles that behave as shift operators, shifting $B$ by the monopole's charge, and also shifting $\sigma$ according to the BPS conditions. Magnetic monopoles have charges labeled by cocharacters ${\rm Hom}(U(1),G)/G \cong {\rm Hom}(U(1), T)/\cW$, where $T\subset G$ is the maximal torus and $\cW$ is the Weyl group of $G$. Thus general nonabelian monopoles have charges labeled by the Weyl group orbits in the lattice of coweights ${\rm Hom}(U(1), T)$. Given such an orbit $\cW b\subset {\rm Hom}(U(1), T)$, where $b$ is one reference coweight (say, the dominant one), the corresponding nonabelian monopole operator can be written as a sum of \emph{abelianized} monopoles $M^{b'}$ over $b'\in\cW b$. For many purposes, $M^{b'}$ behave as monopoles for the gauge group given by the maximal torus $T$. This is, roughly, the core idea of \emph{abelianization}: Solve everything with respect to the maximal torus, and then recover the non-abelian answer by summing over the Weyl orbits. One subtlety this leaves, however, is the possibility of monopole bubbling, which we will address in a moment.

The general expression for the abelianized monopole operator $M^b$ is:
\begin{equation}
M^b = \frac{ \text{Polynomial in }\Phi}{\text{Polynomial in }\Phi} e^{-b\cdot \left(\frac{i}{2}\partial_{\sigma} + \partial_B\right)},
\end{equation}
where polynomials in the numerator and denominator come from the one-loop determinants on the hemisphere, and $e^{-b\cdot \left(\frac{i}{2}\partial_{\sigma} + \partial_B\right)}$ represents the necessary shift of the variables $(\sigma, B)$. The precise expressions are given in \eqref{eq:shift-op} (in fact, there are two representations of these operators, corresponding to the two endpoints of the half-circle $HS^1$, as explained there).

Denominators in the expression for $M^b$ only appear in non-abelian theories, capturing the effects of W-bosons. These denominators also cause some trouble: if we simply take the Weyl-averaged monopole operators $\sum_{b'\in\cW b}M^{b'}$ (as well as the dressed monopoles and Weyl-invariant polynomials in $\Phi$), they will not close into an algebra with polynomial relations. The denominators involving $\Phi$ will appear in relations, which is not good: $\cA_C$ is supposed to be a quantization of the ring $\C[X_C]$ of global functions, and no denominators should be present in the relations. The way out of it comes from addressing another important subtlety.

The BPS equations in a non-abelian theory \cite[eqn. 2.11]{Dedushenko:2018icp} restrict us to $T\subset G$ (i.e., enforce abelianization) almost everywhere. Close to the monopole, however, they look like the Bogomol'ny equations (see \cite{Gomis:2011pf} for the 4D case) and  allow ``solutions at infinity'' (of the filed space), in which dynamical monopoles screen the external singular monopoles \cite{Kapustin:2006pk,Weinberg:1998hn,Houghton:2002bz}, and this screening happens in an infinitesimal neighborhood of the monopole operator. The screening effect is a non-abelian phenomenon somewhat analogous to Nekrasov's instanton partition function \cite{Nekrasov:2002qd} (to which it is related via Kronheimer's correspondence \cite{kronheimer1985monopoles}). However, when it happens in the infinitesimal neighborhood, it violates the ``abelianness'' by an arbitrarily small amount in a way that remains consistent with the SUSY localization. As a result, the correct physical monopoles may be represented not just by the Weyl-average of $M^b$, but rather of $M^b + \sum_v Z_{b\to v}(\Phi)M^v$, where one sums over the possible monopole charges $v$ to which the monopole $b$ can be screened ($v$'s associated to $b$, in the terminology of \cite{Kapustin:2006pk}). Such a screening phenomenon is known as the \emph{monopole bubbling}, and it attracts some interest in the literature. The direct analysis of bubbling contributions is rather hard \cite{Gomis:2011pf,Ito:2011ea,Gang:2012yr,Honda:2017cnz}, and an approach inspired by String Theory was used \cite{Brennan:2018yuj,Brennan:2018moe,Brennan:2018rcn,Assel:2019iae,Brennan:2019hzm}, with further applications (including Coulomb branches) in \cite{Assel:2019yzd,Okuda:2019emk,Hayashi:2020ofu}. For us, the important point is that the \emph{bubbling coefficients} $Z_{b\to v}(\Phi)$ may be rational functions in $\Phi$, so they also have denominators. Moreover, these denominators perfectly conspire in such a way that the combined operators $M^b + \sum_v Z_{b\to v}(\Phi)M^v$, after Weyl-averaging, close into the algebra with only polynomial relations. No denominators appear in the final answer, as is necessary. In fact, we formulate it as a hypothesis, the \emph{polynomiality hypothesis}, which is an assumption under which we work:
\begin{align*}
\boxed{\begin{matrix}\text{$\cA_C$ is generated by Weyl-invariant polynomials $P(\Phi)$ and dressed bubbled monopoles,}\\ \text{subject to polynomial relations.}\end{matrix}}
\end{align*}
The algebraic structure is so tightly knit that this hypothesis alone almost uniquely fixed the bubbling coefficients. We argue that the only freedom in bubbling coefficients not fixed by this hypothesis is the possibility to shift them by polynomials. More generally, it is the freedom of operator mixing, i.e.\ the choice of basis in the space of local operators. We observe empirical evidence for the following \emph{polynomiality conjecture} \cite{Dedushenko:2018icp}:
\begin{align*}
\boxed{\begin{matrix}\text{Polynomiality hypothesis fixes all the bubbling coefficients,}\\ \text{up to the polynomial shifts $Z_{b\to v}(\Phi) \mapsto Z_{b\to v}(\Phi) + P(\Phi)$.}\end{matrix}}
\end{align*}
We can always resolve the operator mixing, so we can equivalently state it as follows:
\begin{align*}
\boxed{\begin{matrix}\text{Assuming the polynomiality hypothesis, we can find a basis of generators,}\\ \text{in which the bubbling coefficients are known.}\end{matrix}}
\end{align*}
This basis may differ from the monopole operators defined via singularity in the path integral, but it allows to fully compute $\cA_C$. The polynomiality conjecture was explicitly proven for theories with full hypermultiplets and (products of) simple gauge groups of rank at most two in \cite{Dedushenko:2018icp}. In the current work, we will extend this to half-hypers and gauge group $SU(2)$ (including products of many copies of $SU(2)$). One curious observation we make is that the preservation of polynomiality hypothesis, i.e., the cancellation of poles, crucially depends on the cancellation of $\Z_2$ anomaly. In the end, we get an explicit set of generators and relations in $\cA_C$, as well as the possibility to compute correlators by applying the gluing formula.

We would like to also mention a few recent papers on Coulomb branches: \cite{Dimofte:2018abu} explicitly mentions the cancellation of poles, \cite{Garner:2023zko} describes spaces of vacua in terms raviolo vertex algebras \cite{Garner:2023zqn}, and \cite{Tamagni:2024ugg} revisits derivation of the BFN Coulomb branch from physics.

The rest of this paper is structured as follows. In Section \ref{sec:3DN4review} we review the structure of 3D $\cN=4$ theories with half-hypermultiplets. We describe them both in $\cN=2$ language and as the real locus of theories with full hypermultiplets, explain the cancellation of $\Z_2$ (parity) anomaly, and provide actions and supersymmetry transformations for theories with half-hypers on $S^3$.

In Section \ref{sec:protected}, we review the Higgs- and Coulomb-branch protected sectors and the results of \cite{Dedushenko:2016jxl,Dedushenko:2017avn,Dedushenko:2018icp}. On the Higgs side we extend the discussion to half-hypermultiplets (this is straightforward), while on the Coulomb side we give additional background for the constructions of \cite{Dedushenko:2017avn,Dedushenko:2018icp}.

In Section \ref{sec:SUSYgluing}, we extend the gluing formula of \cite{Dedushenko:2018aox,Dedushenko:2018tgx} to the case of half-hypers, overcoming some apparent obstacles. In Section \ref{sec:shift-difference}, we describe the hemisphere partition function and the resulting shift/difference operators. We also discuss monopole bubbling and the associated algebraic consistency conditions.

Finally, in Section \ref{sec:examples} we illustrate the formalism in a range of examples.  We first derive generators and relations for $\cA_C$ in $SU(2)$ gauge theory with general matter content (including at least one matter representation of half-integer $SU(2)$ spin).  We then analyze the class of $SU(2)$ theories whose Coulomb branch is $y^2=z(x^2-1)$ (the ``$D_2$'' geometry, which is regular at the origin and has two isolated $A_1$ singularities).  In particular, in an $SU(2)$ theory with a single spin-$\frac32$ half-hypermultiplet, these $A_1$ singularities must support 3D $\cN=4$ SCFTs with $A_1$ Coulomb branch and no Higgs branch.  As a final application we compute the Coulomb branch of an $A_n$ quiver with $SU(2)$ at each node, bifundamental hypers on each edge, and one flavor of spin-$\frac32$ half-hyper attached to each gauge node, as illustrated in Figure \ref{fig:quiver}. After that we conclude with some additional general facts and open problems.

\section{3d $\cN=4$ Theories with Half-Hypermultiplets on $S^3$}\label{sec:3DN4review}

In this section, we review the construction of 3d $\cN=4$ supersymmetric Lagrangians involving half-hypermultiplets coupled to vectormultiplets on $S^3$. The easiest way to construct such theories is to start with full hypermultiplets and impose a reality condition on the hypermultiplet fields, as we elaborate below.

\subsection{General $\cN=4$ Theories and their $\cN=2$ Description}\label{eq:Neq2}

A general 3d $\cN=4$ gauge theory is characterized by the following data: a compact gauge group $G$, a pseudoreal (also called quaternionic) hermitian representation $\mathbf{R}$ of $G$, and a set of mass and Fayet-Iliopoulos (FI) parameters. By pseudoreal representation $\mathbf{R}$ we mean a complex representation that admits a pseudoreal structure $\rho$. The latter is equivalent to having a $G$-invariant symplectic structure on $\mathbf{R}$ that establishes an isomorphism of $\mathbf{R}$ with its dual representation (which is identified with the complex conjugate representation via the invariant hermitian structure),
\begin{align}
\rho:\ &\mathbf{R} \to \overline{\mathbf{R}},\cr 
\bar\rho\circ\rho&=-1,\quad (\text{anti-involution})
\end{align}
where $\bar\rho: \overline{\mathbf R} \to \mathbf{R}$ is the conjugate map. The property $\bar\rho\circ\rho=-1$ prevents us from imposing a reality condition on $\mathbf{R}$ (unlike in the case of real structure). Thus general half-hypers are valued in a symplectic representation $(\mathbf{R},\rho)$ defined by the homomorphism $G\to USp(\mathbf{R})$.

It is often illuminating to view such a theory through the lens of 3d $\cN=2$ subalgebra. From this perspective, the theory consists of:
\begin{itemize}
	\item An $\cN=2$ vectormultiplet with gauge group $G$.
	\item An $\cN=2$ chiral multiplet $\Phi$ transforming in the adjoint representation of $G$.
	\item A set of $\cN=2$ chiral multiplets $q$ transforming in a pseudoreal representation $\mathbf{R}$ of $G$.
	\item A superpotential written in terms of the pseudoreal structure as:
	\begin{align}
	W = \frac12 \rho^{ij}q_j (\Phi q)_i.
	\end{align}
	Here $i,j$ are gauge indices, and $\rho^{ij}$ is the symplectic structure on $\mathbf{R}$ (antisymmetric in $i,j$). This superpotential is necessary for $\cN=4$ supersymmetry.
\end{itemize}
The full hypermultiplet corresponds to $\mathbf{R} = \cR \oplus \overline{\cR}$, where $\cR$ is some complex representation of $G$. In this case,
\begin{equation}
\rho=\left(\begin{matrix}0 & 1\\ -1 & 0\end{matrix}\right)
\end{equation}
is the block-diagonal matrix that swaps $\cR$ wiht $\overline{\cR}$.

The $\cN=2$ vectormultiplet together with the adjoint chiral $\Phi$ form the $\cN=4$ vectormultiplet. We may introduce a triplet of mass parameters by turning on scalar vevs in the background vectormultiplets for flavor symmetries. The vev for scalar $\sigma$ in the background $\cN=2$ vectormultiplet is called real mass, while the vev for background $\Phi$ is called complex mass. Similarly, we may turn on a triplet of FI parameters (associated with background twisted vectormultiplets), whenever the gauge group has non-trivial center.

\subsection{Half-Hypermultiplets from Reality Conditions}\label{sec:reality}

A theory with half-hypermultiplets may also be constructed by starting with full hypermultiplets and imposing the symplectic reality condition. Consider an $\cN=4$ theory with gauge group $G$ and full hypermultiplets $\cH$ transforming in a pseudoreal representation $\mathbf{R}$ of $G$. The components of hypermultiplets are:
\begin{align}
\cH = (q_a, \tq^a, \psi_{\alpha\dot a}, \tpsi_{\alpha\dot a}).
\end{align}
Here $a=1,2$ is the $SU(2)_H$ R-symmetry index, $\dot{a}=1,2$ is the $SU(2)_C$ R-symmetry index, and $\alpha=1,2$ is the spinor index. The scalar fields $q_a$ and fermions $\psi_{\alpha\dot{a}}$ take values in the representation $\mathbf{R}$, while $\tq^a$ and $\tpsi_{\alpha\dot{a}}$  are valued in the complex conjugate (and isomorphic) representation $\overline{\mathbf R}$. In the full hypermultiplet, we only require that $\tq$ is conjugate to $q$:
\begin{equation}
(q_a)^* = \tq^a,
\end{equation}
as well as, in Minkowski signature only, that $\tpsi^{\alpha\dot{a}} = (\psi_{\alpha\dot{a}})^*$. In Euclidean signature, $\tpsi$ and $\psi$ are independent variables. No further reality conditions are needed.

Now let us cut the field content of $\cH$ in half, to construct the half-hypers in $\mathbf{R}$. Let $\rho$ be the invariant antisymmetric form defining the pseudoreal structure on $\mathbf{R}$, satisfying $\rho^2 = -1$. While one cannot impose the reality condition on $\mathbf{R}$ alone, the presence of the R-symmetry index $a$ makes it possible to impose the following additional reality constraint on the bosons:
\es{HalfHyperRealityBosonic}{
	\tq^a = \varepsilon^{ab} \rho q_b,
}
where $\varepsilon^{ab}$ is the antisymmetric tensor with $\varepsilon^{12} = 1$. By supersymmetry, this also forces a condition on the fermionic fields:
\es{HalfHyperRealityFermionic}{
	\tpsi_{\dot a} = \rho \psi_{\dot a}.
}
These conditions are compatible with the SUSY and R-symmetry transformations and halve the number of degrees of freedom, yielding the irreducible half-hypermultiplet.

In the $\cN=2$ language, we take the chiral multiplet $q$ in the complex representation $\mathbf{R}$, and impose that the chiral in $\overline{\mathbf{R}}$ is related to $q$ by $\tq = \rho q$, so the $\cN=4$ superpotential is:
\es{HalfHyperSuperpotential}{
	W = \frac{1}{2} (\rho q) \Phi q.
}
The factor of $1/2$ is included to ensure that when $\mathbf{R} = \cR \oplus \overline{\cR}$, one recovers the standard superpotential $W = \tq \Phi q$ for a full hypermultiplet with chirals $q\in\cR$ and $\tq\in\overline\cR$.

\subsection{The $\Z_2$ anomaly}\label{sec:anomaly}
An important aspect of 3D physics that becomes relevant in the presence of half-hypermultiplets is the parity anomaly \cite{Niemi:1983rq,Redlich:1983kn,Redlich:1983dv}, which we also refer to as the $\Z_2$ anomaly.  The $\Z_2$ anomaly is caused by spinor fields coupled to gauge fields. As a half-hyper consists of a chiral multiplet valued in the symplectic representation $\mathbf{R}$ of $G$, the fermionic kinetic term $\frac{-i}{2}(\rho\psi^{\dot a})\slashed{D}\psi_{\dot a}$, or after integration by parts, $-i (\rho\psi_{\dot 2})\slashed{D}\psi_{\dot 1}$, is simply a Dirac fermion in  $\mathbf{R}$. Integrating it out may produce Chern-Simons (CS) terms. If we give $\psi$ a real mass $M$, it generates a CS level:
\begin{equation}
k_{eff} = \frac12 {\rm Sign}(M) \eta_{\mathbf R}, \quad \text{where the Dynkin index is defined by } \Tr_{\mathbf R}(T^a T^b)=\eta_{\mathbf R} \delta^{ab},
\end{equation}
where we assume the normalization in which $k_{eff}=1$ corresponds to the form $\frac1{2h^\vee}\Tr_{\mathbf{adj}}$ used to define the CS action.

If only gauge fields in the maximal torus $T\subset G$ were activated, for each weight $w$ in $\mathbf{R}$, the corresponding component $\psi^{(w)}_{\dot 1}$ would couple to the gauge field $A^{(w)}\equiv \langle w,A\rangle$. Integrating it out produces the half-integral Chern-Simons term $\frac{{\rm Sign}(M)/2}{4\pi}\int A^{(w)}\dd A^{(w)}$  \cite{Niemi:1983rq,Redlich:1983kn,Redlich:1983dv}. Summing over weights $w$ of $\mathbf{R}$ gives the Weyl-invariant symmetric bilinear form $\tfrac{{\rm Sign}(M)}{2}\sum_w w\otimes w$ on the Cartan subalgebra, whose invariant extension to $\mathfrak{g}$ is $\tfrac{{\rm Sign}(M)}{2} \Tr_{\mathbf R}$, leading to the CS action:
\begin{equation}
\frac{{\rm Sign}(M)/2}{4\pi} \int \Tr_{\mathbf R} \left(A\dd A + \frac23A[A,A]\right),
\end{equation}
which indeed has level $\frac12 {\rm Sign}(M)\eta_{{\mathbf R}}$ in the normalization explained above.

The $\Z_2$ anomaly is the half-integrality, i.e.\ $k_{eff}\mod 1$. It does not depend on ${\rm Sign}(M)$, so the parity anomaly cancellation is the condition that $\eta_{{\mathbf R}}$ is even. It is automatically satisfied for full hypers, since they have ${\mathbf R}=\cR \oplus \overline{\cR}$, and $\eta_{\mathbf R} = 2 \eta_\cR$. In general, the $\Z_2$ anomaly cancellation is the property of form $B$ given by the pullback of $\Tr$ under $\mathfrak{g} \to \mathfrak{sp}(\mathbf{R})$:
\begin{equation}
B(X,Y) = \Tr_{\mathbf R}(XY),
\end{equation}
namely, that it is divisible by 2. It was proven in \cite[Proposition 4.1.2]{Braverman:2022zei} that the $\Z_2$ anomaly cancellation is equivalent to the homomorphism
\begin{equation}\label{wittenZ2}
\pi_4 G \to \pi_4 {\rm Sp}(\mathbf R)
\end{equation}
being trivial. This is also the cancellation condition for Witten's SU(2) anomaly in a 4D theory with chiral matter $\mathbf{R}$ \cite{Witten:1982fp}. In fact, a 4D theory with Witten's $\Z_2$ anomaly reduces to the 3D theory afflicted by the parity anomaly. Both are seen as failure of gauge-invariance under a large gauge transformation. One important difference is that the 4D anomaly renders a theory inconsistent, while the 3D anomaly can be canceled by the CS counterterm. This, of course, breaks parity, hence the name \emph{parity anomaly} (which is a clash between gauge-invariance and parity). We do not care about parity in this paper, so why not just cancel by the CS term? A problem is that a supercymmetric extension of CS term only admits $\cN=3$ SUSY \cite{Kao:1992ig,Gaiotto:2007qi}. Sometimes enhancement to $\cN=4$ happens in the IR \cite{Gaiotto:2008sd}, but generically, it is not expected. Thus, in a generic anomalous theory, we cannot preserve both gauge invariance and $\cN=4$ SUSY (and parity, of course). While one may explore the possibility of SUSY enhancement, we will require manifest $\cN=4$ SUSY, and therefore proceed assuming that the $\Z_2$ anomaly cancels, i.e., that $\eta_{\mathbf R}$ is even (or \eqref{wittenZ2} is trivial).

\subsection{Actions and Supersymmetry on $S^3$}\label{sec:actions}

In practice, it is easy to work directly in the $\cN=4$ component formalism. We borrow the action for full hypermultiplets coupled to vectormultiplets on $S^3$ from \cite{Dedushenko:2016jxl}. To get half-hypers, we substitute the reality conditions \eqref{HalfHyperRealityBosonic} and \eqref{HalfHyperRealityFermionic} into full hypermultiplet action.

The resulting action for gauged half-hypers in the pseudoreal representation $\mathbf{R}$ is:
\begin{align}
S_{\text{half-hyper}}[\cH,\cV] &= \frac12 \int d^3x \sqrt{g} \left[ \rho \cD^{\mu} q^{a} \cD_{\mu} q_a - i (\rho \psi^{\dot{a}})\slashed{D}\psi_{\dot{a}} + \frac{3}{4r^2} (\rho q^{a}) q_a + i (\rho q^{a}) D_a{}^b q_b \right.\cr
&\left.- \frac{1}{2}(\rho q^a) \Phi^{\dot{a}\dot{b}}\Phi_{\dot{a}\dot{b}}q_a - i(\rho \psi^{\dot{a}})\Phi_{\dot{a}}{}^{\dot{b}}\psi_{\dot{b}} +i\left( (\rho q^ a)\lambda_a{}^{\dot{b}}\psi_{\dot{b}} + (\rho \psi^{\dot{a}})\lambda^b{}_{\dot{a}}q_b\right)\right].
\end{align}
Here, the covariant derivative $\cD_\mu$ acts on $q_a$ in the representation $\mathbf{R}$, and $\rho$ acts on the gauge indices. The fermionic path integral for full hypermultiplets on $S^3$ is performed over $\psi_{\dot a}$ and $\tpsi_{\dot a}$ independently. However, after the half-hyper halving \eqref{HalfHyperRealityFermionic}, the remaining fermionic integration is over $\psi_{\dot a}$ alone. Note that we chose to include the factor of $\frac12$ in front of the action to ensure that for $\mathbf{R}=\cR\oplus \overline\cR$, we recover the full hypermultiplet from \cite{Dedushenko:2016jxl}.

The supersymmetry transformations for the half-hypermultiplet are obtained from the full hyper by applying the same halving conditions:
\begin{align}
\delta_{\xi} q^a = \xi^{a\dot{b}} \psi_{\dot{b}},\quad \delta_{\xi} \psi_{\dot{a}} = i\gamma^{\mu}\xi_{a\dot{a}} \cD_{\mu}q^a + i\xi'_{a\dot{a}}q^a - i\xi_{a\dot{c}}\Phi^{\dot{c}}{}_{\dot{a}}q^a.
\end{align}
The transformations for $\tq^a$ and $\tpsi_{\dot a}$ are not independent but are determined by \eqref{HalfHyperRealityBosonic}, \eqref{HalfHyperRealityFermionic}.

The closure of the supersymmetry algebra on the half-hypermultiplet follows from the closure for the full hyper, with the halving conditions being preserved by the algebra. We generally follow the conventions of \cite{Dedushenko:2017avn,Dedushenko:2018icp} where the full hypermultiplet case was analyzed. In particular, the non-conformal supersymmetry algebra on $S^3$, $\mathfrak{su}(2|1)_\ell \oplus \mathfrak{su}(2|1)_r$, and its central extension by real mass and FI parameters, are unchanged from the full hyper case. The SUSY transformations are parameterized in terms of $\xi_{a\dot{a}}$, the conformal Killing spinors on $S^3$, transforming in $(\mathbf{2},\mathbf{2}, \mathbf{2})$ of $SU(2)_H\times SU(2)_C\times SU(2)_E$, where $SU(2)_E = {\rm Spin}(3)$ is the spacetime symmetry, and obeying:
\begin{equation}
\nabla_\mu\xi_{a\dot{a}} = \gamma_\mu \xi'_{a\dot{a}},\quad \nabla_\mu\xi'_{a\dot{a}} = -\frac1{4r^2} \gamma_\mu \xi_{a\dot{a}},
\end{equation}
where $\gamma_\mu = e_\mu^a\sigma_a$, $\mu=1,2,3$, and $\sigma_a$ are Pauli matrices, while $e_\mu^a$ determine the local frame. To be more precise, the solutions to these equations give the full 3D $\cN=4$ superconformal algebra $\mathfrak{osp}(4|4)$ preserved by the free half-hypers on $S^3$, whereas the presence of vectormultiplets and/or mass deformations breaks it down to $\mathfrak{su}(2|1)_\ell \oplus \mathfrak{su}(2|1)_r \subset \mathfrak{osp}(4|4)$. The latter embedding is parameterized in terms of two $2\times 2$ matrices:
\begin{equation}
h_a{}^b \in \mathfrak{su}(2)_H,\quad \bar{h}^{\dot a}{}_{\dot b} \in \mathfrak{su}(2)_C,
\end{equation}
allowing to write the constraint on conformal Killing spinors that determines the non-conformal (or massive) SUSY algebra:
\begin{equation}\label{eq:constraint_hhb}
\xi'_{a\dot{a}} = \frac{i}{2r} h_a{}^b \xi_{b\dot{b}} \bar{h}^{\dot b}{}_{\dot a}.
\end{equation}
The conventions adopted in previous works \cite{Dedushenko:2016jxl,Dedushenko:2017avn,Dedushenko:2018icp} use the following choice of $h, \bar{h}$ in terms of the Pauli matrices:
\begin{equation}
h = - \sigma^2,\quad \bar{h} = -\sigma^3.
\end{equation}

Now we write the vectormultiplet action and the FI term (for a $U(1)$ factor of the gauge group), which are independent of the matter content and thus take the same form as in \cite{Dedushenko:2016jxl}:
\begin{align}
S_{\rm YM}&[\cV] = \frac1{g_{\rm YM}^2} \int \dd^3 x\sqrt{g} \Tr \bigg( F^{\mu\nu}F_{\mu\nu} - \cD^\mu\Phi^{\dot{c}\dot{d}}\cD_\mu\Phi_{\dot{c}\dot{d}} + i\lambda^{a\dot{a}}\slashed{\cD} \lambda_{a\dot{a}} - D^{cd}D_{cd} -i\lambda^{a\dot{a}}[\lambda_a{}^{\dot b}, \Phi_{\dot{a}\dot{b}}]\cr
&-\frac14 [\Phi^{\dot a}{}_{\dot b}, \Phi^{\dot c}{}_{\dot d}][\Phi^{\dot b}{}_{\dot a}, \Phi^{\dot d}{}_{\dot c}] - \frac1{2r} h^{ab}\bar{h}^{\dot{a}\dot{b}}\lambda_{a\dot{a}}\lambda_{b\dot{b}} + \frac1{r} (h_a{}^b D_b{}^a)(\bar{h}^{\dot a}{}_{\dot b}\Phi^{\dot b}{}_{\dot a}) - \frac1{r^2}\Phi^{\dot{c}\dot{d}}\Phi_{\dot{c}\dot{d}}\bigg),\\
S_{\rm FI}&[\cV] = i\zeta \int \dd^3x \sqrt{g} \left(h_a{}^b D_b{}^a -\frac1{r} \bar{h}^{\dot a}{}_{\dot b}\Phi^{\dot b}{}_{\dot a}\right).
\end{align}
The vectormultiplet SUSY transformations are:
\begin{align}
\delta_{\xi} A_{\mu} &= \frac{i}{2} \xi^{a\dot{b}}\gamma_{\mu}\lambda_{a\dot{b}} \ec \label{Avar}\\
\delta_{\xi} \lambda_{a\dot{b}} &= - \frac{i}{2}\varepsilon^{\mu\nu\rho}\gamma_{\rho}\xi_{a\dot{b}}F_{\mu\nu} - D_a{}^c\xi_{c\dot{b}} -i\gamma^{\mu}\xi_a{}^{\dot{c}} \cD_{\mu}\Phi_{\dot{c}\dot{b}} + 2i \Phi^{\dot{c}}{}_{\dot{b}}\xi_{a\dot{c}}' \notag\\
&+ \frac{i}{2}\xi_{a\dot{d}} [ \Phi_{\dot{b}}{}^{\dot{c}}, \Phi_{\dot{c}}{}^{\dot{d}}] \ec \label{lamvar}\\
\delta_{\xi}\Phi_{\dot{a}\dot{b}} &= \xi^c{}_{(\dot{a}}\lambda_{|c|\dot{b})} \ec \label{phivar}\\
\delta_{\xi} D_{ab} &= -i\cD_{\mu}(\xi_{(a}{}^{\dot c}\gamma^{\mu}\lambda_{b)\dot c}) - 2i\xi'_{(a}{}^{\dot c}\lambda_{b)\dot c} + i [\xi_{(a}{}^{\dot{c}}\lambda_{b)}{}^{\dot{d}}, \Phi_{\dot{c}\dot{d}}] \ed\label{dvar}
\end{align}

The flavor symmetry group is given by the normalizer of the image of the gauge group $G$ inside the symplectic group associated with the pseudoreal representation $G \to Sp(\mathbf{R})$:
\begin{equation}
G_f = \cN_G(Sp(\mathbf{R})).
\end{equation}
Finally, we can turn on the background vectormultiplet $\cV_{\rm b.g.}$ in the Cartan $\mathfrak{t}_f \subset \mathfrak{g}_f$ of the flavor group, with the constant background values subject to the following constraint:
\begin{equation}
2i (\Phi_{\rm b.g.})^{\dot{c}}{}_{\dot{b}}\xi_{a\dot{c}}' - (D_{\rm b.g.})_a{}^c\xi_{c\dot{b}}=0,
\end{equation}
ensuring $\delta_\xi\lambda_{\rm b.g.}=0$. Using \eqref{eq:constraint_hhb}, this is written as:
\begin{equation}
\frac{1}{r} (\Phi_{\rm b.g.})^{\dot{c}}{}_{\dot{b}} h_a{}^b \xi_{b\dot{a}} \bar{h}^{\dot a}{}_{\dot c} + (D_{\rm b.g.})_a{}^c\xi_{c\dot{b}}=0,
\end{equation}
which has a unique solution in terms of $m\in \mathfrak{t}_f$:
\begin{equation}
(\Phi_{\rm b.g.})^{\dot{a}}{}_{\dot{b}} = -m\, \bar{h}^{\dot a}{}_{\dot b},\quad (D_{\rm b.g.})_a{}^b = \frac{m}{r} h_a{}^b.
\end{equation}
This gives the real mass deformations.\footnote{In flat space, $SU(2)_C$-triplet of masses and $SU(2)_H$-triplet of FI parameters exist, but only one linear combination of masses/FI parameters, which we call the \emph{real mass/FI parameter}, is allowed on $S^3$.} In our conventions, this gives $(\Phi_{\rm b.g.})_{\dot{1}\dot{2}}=m$.

\section{Protected sectors}\label{sec:protected}
As studied in \cite{Dedushenko:2016jxl,Dedushenko:2017avn,Dedushenko:2018icp}, special supercharges $\cQ_H, \cQ_C \in \mathfrak{su}(2|1)_\ell\oplus \mathfrak{su}(2|1)_r$ define the Higgs and Coulomb branch ``protected sectors''. Namely, supersymmetric localization with respect to $\cQ_H$ (respectively, $\cQ_C$) computes correlation functions of the Higgs (respectively, Coulomb) branch operators in the cohomology of $\cQ_H$ (respectively, $\cQ_C$). The data of such correlation functions constitutes the protected sector encoded in an algebra $\cA_H$ (respectively, $\cA_C$), equipped with a twisted trace. The algebras $\cA_H$ and $\cA_C$ are short quantizations \cite{Etingof:2019guc} of the Higgs and Coulomb branches (later called ``sphere quantization'' \cite{Gaiotto:2019mmf}).

More precisely, we need \emph{equivariant} cohomology, because $\cQ_H$ and $\cQ_C$ are not nilpotent:
\begin{align}
(\cQ_H)^2 &= \frac{i}{r} (P_\tau + R_C + ir\widehat\zeta), \cr
(\cQ_C)^2 &= \frac{i}{r} (P_\tau + R_H + ir\widehat{m}),
\end{align}
where we work in the spherical coordinates $(\theta, \tau, \varphi)$ on $S^3$ used in \cite{Dedushenko:2016jxl}, and $P_\tau$ is the operator that rotates the $\tau$ angle. Here $R_H=\frac12 h_a{}^b R_b{}^a$ and $R_C=\frac12 \bar{h}^{\dot a}{}_{\dot b} \bar{R}^{\dot b}{}_{\dot a}$ are the Cartan generators of $SU(2)_H$ and $SU(2)_C$, related to the generators of $\mathfrak{u}(1)_\ell\oplus \mathfrak{u}(1)_r\subset \mathfrak{su}(2|1)_\ell\oplus \mathfrak{su}(2|1)_r$ via $R_\ell=R_H+R_C$ and $R_r=R_H-R_C$. Finally, $\widehat\zeta$ and $\widehat{m}$ are FI and mass central charges, which act on states charged under the topological and flavor symmetries, respectively. The local operators in the equivariant cohomology of $\cQ_H$ must be annihilated by $\cQ_H^2$, which is achieved by placing them at the fixed point locus of $\partial_\tau$ and making sure that they are annihilated by $R_C + ir\widehat\zeta$ (which is true for the Higgs branch operators). Same is true for $\cQ_C$, in which case we are dealing with the Coulomb branch operators annihilated by $R_H + ir\widehat{m}$.

The fixed point locus of $\partial_\tau$ is a great circle $S^1_\varphi\subset S^3$ parameterized by $\varphi$. The cohomology of $\cQ_H$ and $\cQ_C$ (on local observables) are spanned by the ``twisted-translated'' local operators inserted along $S^1_\varphi$: The twisted-translated Higgs branch operators for $\cQ_H$, and the twisted-translated Coulomb branch operators for $\cQ_C$.

\subsection{The Higgs sector}
The twisted-translated (half-)hypermultiplet letters are:
\begin{align}
Q(\varphi) = q_1 \cos\frac{\varphi}{2} + q_2 \sin\frac{\varphi}{2},\quad \tQ(\varphi) = \tq_1 \cos\frac{\varphi}{2} + \tq_2 \sin\frac{\varphi}{2},
\end{align}
and general twisted-translated Higgs branch operators are constructed as gauge-invariant words built from these letters. In the case of half-hypers described as hypers in $\mathbf{R}$ subject to the reality condition, we have $\tq_1=-\tq^2=\rho q_1$ and $\tq_2=\tq^1= \rho q_2$, hence 
\begin{equation}
\tQ = \rho Q,
\end{equation}
and we only have one independent variable, the $\mathbf{R}$-valued scalar $Q$.

After the supersymmetric localization using $\cQ_H$, the effective 1D description of correlators along $S^1_\varphi$ emerges. In \cite{Dedushenko:2016jxl} this problem was fully solved for the full hypermultiplets, where a gauged topological quantum mechanics was identified as the exact 1D description. Equivalently, it is a topological quantum mechanics coupled to the matrix model, with the matrix variable $\sigma\in\mathfrak{t}\subset \mathfrak{g}$. That answer extends to the half-hypermultiplets in the obvious way, and here is why.

The approach in \cite{Dedushenko:2016jxl} was to first localize the vectormultiplets, and then the hypers. After the first step, one is left with an abelian vectormultiplet background labeled by $\sigma \in \mathfrak{t} \subset \mathfrak{g}$ (that is integrated over) and a one-loop determinant of the vectormultiplets. After that, the half-hypers are only coupled to the abelian vectormultiplet in the maximal torus $T\subset G$.

Notably, the abelian group $T= U(1)^r$ does not admit pseudoreal representations. In fact, the pseudoreal representation $\mathbf{R}$ of $G$, after restricting to $T$, breaks into a sum of one-dimensional complex representations. (For example, the $\mathbf{2}$ of $SU(2)$ becomes $\C_{+1}\oplus \C_{-1}$ as a representation of $U(1)\subset SU(2)$.) The half-hypermultiplet in $\mathbf{R}$ becomes a collection of abelian full hypermultiplets when the gauge group is restricted to $T\subset G$. Thus, after the vectormultiplet localization, the next step is to localize a bunch of full hypermultiplets coupled to the abelian gauge fields. This was already done in \cite{Dedushenko:2016jxl}, so we have the answer. The 1D sector is described by the path integral of gauged quantum mechanics:
\begin{equation}
S_{\rm 1D} =2\pi r \int \dd\varphi\, Q \rho \cD_\varphi Q,
\end{equation}
or after gauge-fixing the 1D gauge field, one has the equivalent form of the answer as:
\begin{equation}
Z = \frac1{|\cW|}\int_{\mathfrak t} \dd\sigma \det{}'_{\rm adj} \left(2\sinh \pi\sigma\right) \int DQ e^{-S_\sigma[Q]},
\end{equation}
where
\begin{equation}\label{eq:topQM}
S_\sigma[Q] = 4\pi r\int\dd\varphi \left[\tfrac12 Q \rho\, \partial_\varphi Q + \tfrac12 Q \rho (\sigma + mr)Q - 2\pi i \zeta\cdot\sigma\right],
\end{equation}
and we included the contributions of masses and FI terms.

Thus the answer for the 1D theory is, basically, identical to the one derived in \cite{Dedushenko:2016jxl}, except we let $Q$ be valued in $\mathbf{R}$ and take $\tQ = \rho\, Q$, instead of it being an independent variable. The symplectic structure $\rho$ of $\mathbf{R}$ thus naturally becomes the symplectic structure in the first order action \eqref{eq:topQM}.

This answer can then be used to study the Higgs sector, which is not our main goal here.

\subsection{The Coulomb sector}
The Coulomb sector description given in \cite{Dedushenko:2017avn,Dedushenko:2018icp} for the case of full hypers is very different. Instead of an explicit 1D action, the answer is formulated in terms of a certain algebra of difference (or shift) operators, its module, and an inner product on the module. More precisely, the algebra $\cA_C$ itself (the ``quantized Coulomb branch'') is simply the algebra of such difference operators, whereas the twisted trace on $\cA_C$ (that allows to compute correlation functions) is encoded in the structure of module with the inner product.

Let us briefly recall the details. Firstly, the operators in the $\cQ_C$-cohomology are Coulomb branch chiral ring operators that undergo a $U(1)_C$ rotation as we travel around the $S^1$. They are constructed from two building blocks: the twisted-translated vector multiplet scalar,
\begin{equation}
\Phi(\varphi) = \Phi_{\dot{a}\dot{b}} v^{\dot a} v^{\dot b},\quad v = \frac1{\sqrt 2} \left(\begin{matrix}e^{i\varphi/2}\\ e^{-i\varphi/2}\end{matrix}\right),
\end{equation}
and the twisted-translated BPS monopole operator $M_b(\varphi)$, a point-like source of magnetic flux. These are the GNO monopoles \cite{Goddard:1976qe} constructed by embedding the singular Dirac's $U(1)$ monopole into the non-abelian group $G$. Such embeddings are labeled by the cocharacters ${\rm Hom}(U(1), G)/G \cong {\rm Hom}(U(1), \mathbb{T})/\cW$, or the GNO charges. An element of ${\rm Hom}(U(1), \mathbb{T})/\cW$,  seen as a map of algebras $\R \to \mathfrak{t}$, sends $1\mapsto b$ subject to $e^{2\pi b}=1$ (so $b$ is a coweight), up to the action of Weyl group. So cocharacters correspond to $\cW$-orbits in the lattice of coweights $\Lambda^\vee \subset \mathfrak{t}$. A bare monopole operator associated to a given cocharacter is a sum over the corresponding Weyl orbit, in which a summand labeled by the coweight $b\in\Lambda^\vee$ from the $\cW$-orbit is a local disorder operator characterized by the following singular behavior of the gauge and scalar fields:
\begin{equation}\label{eq:monopole_singularity}
\star F \sim b \frac{y_\mu \dd y^\mu}{|y|^3},\quad \Phi_{\dot{1}\dot{1}} = - (\Phi_{\dot{2}\dot{2}})^\dagger \sim -\frac{b}{2|y|} e^{-i\varphi},\quad \Phi_{\dot{1}\dot{2}} \sim 0.
\end{equation}
Here $y^\mu$ are Riemann normal coordinates centered at the monopole. The singular behavior of scalars in \eqref{eq:monopole_singularity} is, of course, dictated by supersymmetry \cite{Borokhov:2002cg}, and the explicit factor of $e^{-i\varphi}$ reflects the name ``twisted-translated'' \cite{Dedushenko:2017avn}. 

More generally, the monopole may be dressed by some polynomial in $\Phi(\varphi)$, subject to the appropriate gauge-invariance condition. The monopole singularity breaks the gauge group at the insertion point down to the subgroup $G_b \subset G$, so the dressing factor $P(\Phi)$ must only be $G_b$-invariant. Thus general Coulomb branch operators are dressed monopoles (including those of GNO charge $0$, which are just ${\rm ad}(G)$-invariant polynomials in $\Phi$).

We see that the construction of Coulomb branch operators only depends on the gauge group $G$, not the matter content. We are interested in their correlation functions on $S^1_\varphi\subset S^3$, which, of course, depend on the matter content. The solution to this problem in the case of full hypermultiplet matter was given in \cite{Dedushenko:2017avn,Dedushenko:2018icp}, where a general Coulomb branch operator is represented by
\begin{equation}
\sum_{v\leq b}\sum_{{\rm w} \in \cW} P(\Phi^{\rm w}) M^{{\rm w}\cdot v},
\end{equation}
where we pick a dominant coweight $b$ (each Weyl orbit contains precisely one such coweight), as well as its ``associated'' coweights $v< b$ corresponding to the \emph{bubbling phenomenon} \cite{Kapustin:2006pk} (``associated'' means that $v$ appears in the highest-weight irrep of highest weight $b$), and for each $v\leq b$, we sum over its Weyl orbit, with the summand being the dressed monopole $P(\Phi^{\rm w}) M^{{\rm w}\cdot v}$, where $M^{{\rm w}\cdot v}$ is determined by the singularity \eqref{eq:monopole_singularity} (with $b$ replaced by ${\rm w}\cdot v$). Here $M^v$ is a certain difference operator given in \cite{Dedushenko:2018icp}, which we do not write yet, but it will be given later in this paper for the case of half-hypermultiplet matter. As for $\Phi$, it was represented as a multiplication operator, such as $\Phi = \tfrac1r \left(\sigma + \tfrac{i}{2}B\right)$, acting on functions of coweight $B\in \Lambda^\vee$ and $\sigma\in \mathfrak{t}$. The latter functions, in fact, form a module on which the algebra $\cA_C$ of difference operators acts. We think of elements of this module as wavefunctions in variables $(\sigma, B)$ equipped with the inner product (the ``gluing formula''):
\begin{equation}\label{eq:inner-prod}
(f, g) = \frac1{|\cW|}\sum_{B\in \Lambda^\vee} \int_{\mathfrak t} \dd^r\sigma\, \mu(\sigma,B) f(\sigma, B) g(\sigma, B),
\end{equation}
where $\mu(\sigma, B)$ is a certain ``gluing'' measure. Again, we do not give a formula for $\mu(\sigma, B)$ now, since we will derive it later for the case of general matter.

\begin{figure}[h]
	\centering
	\includegraphics[width=0.3\textwidth]{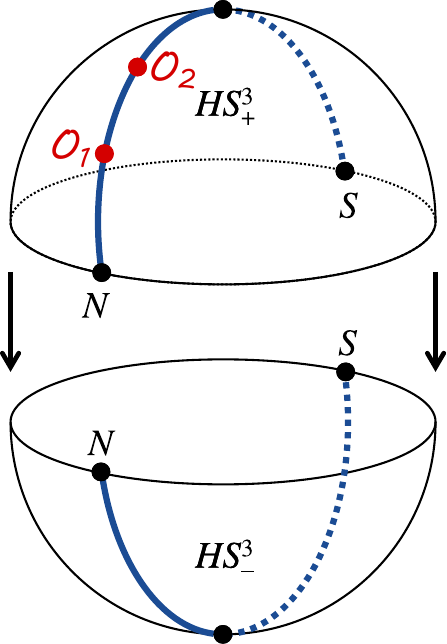}
	\caption{Each hemisphere partition function is viewed as a ``wavefunction'', and the gluing operation -- as an inner product \eqref{eq:inner-prod}. The insertions of twisted-translated Coulomb branch operators are represented via certain shift operators acting on the hemisphere wavefunction.}\label{fig:gluing-sketch}
\end{figure}

An important ingredient in \cite{Dedushenko:2017avn,Dedushenko:2018icp} was this gluing formula, which identifies the $S^3$ partition function as the inner product of two hemisphere partition functions. Each hemisphere $HS^3$ was represented by certain ``wavefunction'' $\psi(\sigma, B)$, equal to the hemisphere partition function with special boundary conditions labeled by $\sigma, B$. The insertions of monopole operators inside the hemisphere were represented by the difference/shift operators sketched earlier acting on the hemisphere partition function $\psi(\sigma, B)$. This procedure is depicted on the Figure \ref{fig:gluing-sketch}, and in the following chapters, we will elaborate and generalize it in the presence of half-hypermultiplet matter.

\section{Supersymmetric Gluing Formula}\label{sec:SUSYgluing}
\subsection{Why the old approach requires an upgrade}\label{sec:OldGluing}
We would like to generalize the gluing formula of \cite{Dedushenko:2017avn,Dedushenko:2018icp}, expressing the $S^3$ partition function as an inner product of two hemispheres. The previous approach was to zoom in on the infinitesimal neighborhood of the equator $S^2\subset S^3$, where we approximately view the geometry as a local segment of $S^2\times \R$. To $S^2$, one associates the field-theoretic phase space $\cP[S^2]$, which is an infinite-dimensional space in which the classical time evolution is described by the first order equations (Hamiltonian equations of motion). Quantum-mechanically, this formally leads to the infinite-dimensional version of Schrodinger equation on ``wave functionals'', which only depend on half of the canonical coordinates/momenta that Poisson-commute. The gluing in \cite{Dedushenko:2018aox,Dedushenko:2018tgx} is identified as an integral over such a half of variables, which can be seen as an integral over the Lagrangian submanifold of $\cP[S^2]$. It is of course infinite-dimensional and looks like a path integral for a 2D theory along $S^2$. By carefully choosing the Lagrangian submanifold in $\cP[S^2]$, one could ensure that this 2D path integral had $\cN=(2,2)$ SUSY, which then afforded the application of localization techniques.

For 3D $\cN=4$ vector multiplet, the Lagrangian submanifold in $\cP[S^2]$ was chosen such that the integral over it looked like an integral over a 2D $\cN=(2,2)$ vector multiplet on $S^2$. Basically, the 2D variables were the pullback of the 3D gauge field to $S^2$ and its 2D $\cN=(2,2)$ SUSY completion. For 3D $\cN=4$ hypermultiplets, however, the preservation of 2D $\cN=(2,2)$ SUSY in the gluing integral required a choice of holomorphic-Lagrangian splitting of the matter representation. For the full hypermultiplets, the splitting
\begin{equation}
\mathbf{R} = \cR \oplus \bar\cR
\end{equation}
already does the job, as it is holomorphic-Lagrangian in the complex structure $I$ (in which the $\cN=2$ chiral multiplets are valued in $\cR$ and $\overline\cR$,) and, obviously, gauge-invariant. With such a choice, we had a well-defined $\cN=(2,2)$ supersymmetric gluing.

For half-hypers, however, no such gauge-invariant Lagrangian splitting exists. Indeed, if we were able to find a $G$-invariant splitting $\mathbf{R} = \mathbb{L} \oplus \mathbb{L}^\perp$, we would simply get full hypermultiplets in $\mathbb{L}$. This poses the main challenge of the generalization, and we must come up with a different method to circumvent this challenge. The vectormultiplet gluing (if we somehow separate the gauge sector from the matter sector) is not affected by this problem, since its form does not depend on the matter content. Thus, roughly, we want to first perform the gluing for the vectormultiplets, and then deal with the half-hypers. The vectormultiplet gluing involves gauge-fixing along the $S^2$, after which the lack of gauge-invariant splitting $\mathbf{R} = \mathbb{L} \oplus \mathbb{L}^\perp$ is no longer an issue. After fixing gauge, the splitting no longer must be gauge-invariant, and we can pick a non-gauge-invariant one. Now we would like to make this argument more precise.

\subsection{The cutting interface}\label{sec:CuttingInterface}

We would like to add a supersymmetric interface on the equator $S^{2}\subset S^{3}$ that implements the cutting-and-gluing. We do it for the gauge sector first, and then include the matter.
We continue using spherical coordinates $(\theta,\tau,\varphi)$ on $S^{3}$, with the equator at $\theta=\pi/2$, and regard $(\tau,\varphi)$ as coordinates on $S^{2}$.

Like in \cite{Dedushenko:2017avn}, we identify a $2$D $\cN=(2,2)$ \emph{vector} multiplet by restricting the $3$D $\cN=4$ vectormultiplet to the equator $S^2$ and identify the submultiplet closed under the $\cN=(2,2)$ subalgebra\footnote{While circle reduction preserves all SUSY, restriction to a subspace can at most preserve half of SUSY.}. A $Q$-exact Yang–Mills term built solely from the fields of this $2$D vector multiplet localizes the equatorial degrees of freedom and enforces the matching of vectormultiplet data across the two hemispheres.

In flat space, the choice of 2D $\cN=(2,2)$ subalgebra is associated with the choice of vector and axial R-symmetry, $U(1)_V\times U(1)_A \subset SU(2)_H \times SU(2)_C$. The latter choice is continuous, parameterized by $\frac{SU(2)_H}{U(1)_H} \times \frac{SU(2)_C}{U(1)_C}$, up to an additional $\Z_2$ that determines whether $U(1)_H$ or $U(1)_C$ is the vector R-symmetry. In particular, the sphere $SU(2)_C/U(1)_C$ (the twistor sphere of the Coulomb branch) determines which linear combination of scalars $\Phi_{\dot{a}\dot{b}}$ is chosen as a real scalar $\sigma$, and which -- as a complex scalar $\phi$. Associated with such a choice, one can split the 3D $\cN=4$ vector multiplet in two ways. One is the familiar 3D $\cN=2$ decomposition into a vector $(A_\mu,\sigma,\ldots)$ and an adjoint chiral multiplet $(\phi,\ldots)$. Another is a decomposition in terms of 2D $\cN=(2,2)$ multiplets, splitting the 3D coordinates (in Euclidean signature) as $(x^{i=1,2}, x^\perp)$. In this language, there is a pair $(V,S)$ of a $(2,2)$ vectormultiplet and a $(2,2)$ adjoint chiral multipet on $\R^2_{x^1, x^2}$, with the infinite-dimensional gauge group ${\rm Maps}(\R_{x^\perp}, G)$, as explained in \cite[Appendix A]{Bullimore:2016nji}. Here $V$ contains the 2D gauge field $A_i$ and the complex scalar $\phi$, while $S$ starts with $A_\perp + i\sigma$. These multiplets are especially convenient for constructing $(2,2)$ boundary conditions at $x^\perp=0$, since $V$ and $S$ each close under the off-shell 2D $\cN=(2,2)$ SUSY, and for each variable in $V$, the canonically conjugate one  belongs to $S$. In \cite[Appendix A]{Bullimore:2016nji}, these multiplets were identified in flat space.

On $S^3$, the $SU(2)_H \times SU(2)_C$ is already broken to $U(1)_H\times U(1)_C$, so the identification of $\cN=(2,2)$ along the equator involves no continuous choices. Up to automorphisms, only a discrete ($\Z_2$) choice is left, leading to the two subalgebras known as:
\begin{equation}
\mathfrak{su}(2|1)_A \subset \mathfrak{su}(2|1)_\ell\oplus \mathfrak{su}(2|1)_r,\quad \mathfrak{su}(2|1)_B \subset \mathfrak{su}(2|1)_\ell\oplus \mathfrak{su}(2|1)_r,
\end{equation}
where $\mathfrak{su}(2|1)_A$ contains $U(1)_H$ and $\mathfrak{su}(2|1)_B$ has $U(1)_C$ as the R-symmetry.

\subsubsection*{Field content on the interface}

Like in \cite{Dedushenko:2017avn}, we choose to work with the $\mathfrak{su}(2|1)_A$ subalgebra, since it contains $\cQ_C$, and the corresponding Coulomb branch localization formula in 2D \cite{Doroud:2012xw,Benini:2012ui} is well-suited for our purposes. With our conventions, $\frac{\Phi_{\dot1\dot1}-\Phi_{\dot2\dot2}}{2}$ is the real scalar and $\frac{\Phi_{\dot1\dot1}+\Phi_{\dot2\dot2}}{2i} - i\Phi_{\dot1\dot2}$ -- the complex scalar. The precise identification of multiplets on the equator $S^2\subset S^3$ is as follows:

\begin{itemize}
	\item A \textbf{$2$D $\cN=(2,2)$ vector multiplet} contains fields
	\[
	(A_i,\; \phi,\; \lambda_\pm,\; \widetilde\lambda_\pm,\; D), \qquad i=1,2,
	\]
	where $A_i$ is the pullback of the gauge field to $S^2$. The complex scalar $\phi$, the auxiliary field $D$, and the fermions $\lambda, \widetilde\lambda$ are expressed in terms of the 3D fields as:
	\begin{align}
	\phi &= \phi_1 + i \phi_2,\quad \phi_1 = \frac{\Phi_{\dot{1}\dot{1}} + \Phi_{\dot{2}\dot{2}}}{2i},\quad \phi_2 = -\Phi_{\dot{1}\dot{2}},\\
	D &= -\frac{\Phi_{\dot{1}\dot{2}}}{r} + \frac{i}{2} (D^{\rm on}_{11} + D^{\rm on}_{22}) + i\cD_\perp \frac{\Phi_{\dot{1}\dot{1}} - \Phi_{\dot{2}\dot{2}}}{2},\\
	\lambda &= -\frac12 (\lambda_{1\dot{2}} - i\lambda_{2\dot{2}} + \sigma_3(\lambda_{1\dot{1}} - i\lambda_{2\dot{1}})),\\
	\widetilde\lambda &= -\frac12 (\lambda_{1\dot{2}} + i\lambda_{2\dot{2}} - \sigma_3(\lambda_{1\dot{1}} +i \lambda_{2\dot{1}})).
	\end{align}
	Here $D^{\rm on}_{ab}$ denotes the on-shell value of the 3D auxiliary field, i.e., we assume that $D_{ab}$ has been integrated out in the bulk. This on-shell value is:
	\begin{equation}
	\frac{i}2 (D_{11}^{\rm on}+D_{22}^{\rm on})= \frac{\Phi_{\dot1\dot2}}{r} + \frac{g^2}{8} \left[\rho q_1 T^A q_1 + \rho q_2 T^A q_2\right] - i\zeta \frac{g^2}{2},
	\end{equation}
	so that the 2D auxiliary field is:
	\begin{equation}
	D = \frac{i g^2}{4} \Im \left[\rho q_2 T^A q_2\right] - i \zeta \frac{g^2}{2} + i\cD_\perp \frac{\Phi_{\dot1\dot1}-\Phi_{\dot2\dot2}}{2}.
	\end{equation}
	\item A \textbf{$2$d $\cN=(2,2)$ adjoint chiral multiplet} with scalar $A_\perp + i\frac{\Phi_{\dot1\dot1}-\Phi_{\dot2\dot2}}{2}$, whose precise form is not relevant.
\end{itemize}

\subsubsection*{Interface-localization term}

We then proceed to add a localizing term along the equator,
\begin{equation}
e^{-t \int_{S^2}\dd^2 x\, \cQ_C V},
\end{equation}
where $t\to\infty$ restricts us to the localization locus. Schematically, if we have a field $A$ in the bulk (such as the gauge field), denoting its equator restriction as $a=A\big|_{S^2}$, we may write the functional integration over $A$ on $S^3$ as follows:
\begin{equation}
\int \cD A (\dots) = \int \cD a \int_{A\big| = a} \cD A (\dots), 
\end{equation}
where the outer integral is over the filed $a$ on $S^2$, and the inner integral is over $A$ on $S^3$ subject to the condition $A\big|=a$. Now as we perform the localization via a term $e^{-t \int_{S_2} L_{\rm loc}(a, \dd a, \dots)}$ on $S^2$, we write $a = a_0 + \frac1{\sqrt{t}} b$, where $a_0$ is on the localization locus. This is useful when the localization locus is finite-dimensional, so the integral over $a_0$ is ordinary, not functional. This leads to the following approximation:
\begin{equation}
\int \cD a \int_{A\big| = a} \cD A (\dots) \Rightarrow \int \dd a_0 \int \cD b e^{- \int_{S^2}\dd^2 x\, b\Delta b} \int_{A\big| = a_0 + \frac1{\sqrt{t}}} \cD A (\dots),
\end{equation}
which becomes exact in the $t\to +\infty$ limit. Here $\Delta$ is the general operator describing quadratic action for the fluctuation field $b$. In the end we obtain:
\begin{equation}
\int \dd a_0 \int \cD b e^{- \int_{S^2}\dd^2 x\, b\Delta b} \int_{A\big| = a_0 + \frac1{\sqrt{t}}} \cD A (\dots) \xrightarrow{t\rightarrow +\infty} \int \dd a_0 \frac1{\sqrt{\det \Delta}} \int_{A\big| = a_0} \cD A (\dots).
\end{equation}
After these steps, we essentially have performed the cutting and gluing: the answer looks like two hemisphere path integrals subject to $A\big|=a_0$, glued together via the $a_0$ integration, with the one-loop determinant on $S^2$ playing the role of gluing measure.

This is the procedure of \cite{Dedushenko:2018tgx} reformulated in a slightly different way. Let us first do such cutting-and-gluing for the vectormultiplets, using the 2D $\cN=(2,2)$ vector multiplet identified above. We apply the Coulomb branch localization formula of \cite{Doroud:2012xw,Benini:2012ui}, just like was done for the case of full hypers in \cite{Dedushenko:2017avn}. The path integral over 2D $\cN=(2,2)$ vectors localizes (after gauge-fixing) to the integral over the Cartan subalgebra $\mathfrak{t}\subset \mathfrak{g}$ and the sum over the coweights $B\in \Lambda^\vee$ (magnetic fluxes through $S^2$), leading to the following presentation of the 3D path integral:
\begin{equation}
\sum_{B\in\Lambda^\vee} \frac1{|\cW(H_B)|} \int_{\mathfrak{t}} \dd^r\sigma\, Z_{\rm vm}(\sigma, B) \int_{\text{conditions on }S^2} \cD [\text{3D fields}] (\dots),
\end{equation}
where $\cW(H_B)$ is the Weyl group of the subgroup $H_B\subset G$ preserved by the flux $B$,
the vectormultiplet one-loop determinant on $S^2$ is a product over positive roots:
\begin{equation}
Z_{\rm vm}(\sigma, B) = \prod_{\alpha\in \Delta^+} (-1)^{\alpha\cdot B} \left[(\alpha\cdot B/2)^2 + (\alpha\cdot\sigma)^2\right],
\end{equation}
and the integral over 3D fields subject to the ``conditions on $S^3$'' means the same set of boundary condition as in \cite{Dedushenko:2017avn,Dedushenko:2018icp}:
\begin{align}\label{eq:VM-bc}
A\big|_{S^2} = \frac{B}{2} (\sin\theta - 1)\dd\tau,\quad \frac1{2i}(\Phi_{\dot1\dot1} + \Phi_{\dot2\dot2})\big|_{S^2} = \frac{B}{2r},\quad \Phi_{\dot1\dot2}\big|_{S^2} = \frac{\sigma}{r},\quad D=0,\cr
(\lambda_{1\dot2} - i \sigma_3 \lambda_{2\dot1})\big|_{S^2}=0,\quad (\lambda_{2\dot2} + i\sigma_3 \lambda_{1\dot1})\big|_{S^2}=0.
\end{align}
Crucially, after this step, the gauge freedom is completely fixed along $S^2$. The remaining gauge transformations are constant along $S^2$, i.e., for each $(\sigma, B)$ the gauge symmetry becomes global symmetry $G_\partial$ along $S^2$, where $G_\partial$ is the centralizer of $\sigma$ and $B$. Generic $\sigma, B$ will break $G$ to the maximal torus, so $G_\partial = T$, while at $\sigma=B=0$, the full group is restored, but only as a global symmetry $G_\partial=G$.

Now we can address the half-hypermultiplet matter. Since we no longer have any gauge redundancies on $S^2$, nothing stops us from splitting the half-hypermultiplets (restricted to $S^2$) into the holomorphic Lagrangian subspaces:
\begin{equation}
\mathbf{R} = \mathbb{L} \oplus \mathbb{L}^\perp.
\end{equation}
More specifically, we think of $\mathbb{L}$ and $\mathbb{L}^\perp$ as dual weight subspaces for the maximal torus $T$. Indeed, $(B, \sigma)$ give an abelian background, so we may restrict the pseudoreal representation $\mathbf{R}$ of $G$ to the maximal torus $T$. This turns half-hypermultiplets into full hypermultiplets with respect to the abelian group $T$.

\paragraph{Example.} The simplest example of the above phenomenon can be seen for the fundamental of $SU(2)$, which is a pseudoreal representation. A fundamental half-hypermultiplet contains two scalar fields, corresponding to the weights $\pm\frac12$ of $SU(2)$. After restriction to the maximal torus $U(1)\subset SU(2)$, these two weights become two separate representations of $U(1)$ of charges $\pm1$, forming a full hypermultiplet with respect to $U(1)$.

Thus we can perform the cutting-and-gluing (through the similar insertion of a localizing interface on $S^2\subset S^3$) for abelian hypermultiplets charged in $\mathbb{L}$. This leads to the following boundary conditions along $S^2$ derived in \cite{Dedushenko:2017avn}:
\begin{align}\label{eq:HM-bc}
q^{\mathbb{L}}_+\big|_{S^2}=0,\quad \left(\cD_\perp q^{\mathbb{L}}_- + \frac12(\Phi_{\dot1\dot1}-\Phi_{\dot2\dot2})q^{\mathbb{L}}_-\right)\big|=0,\quad (\psi^{\mathbb{L}}_{\dot1} - \sigma_3\psi^{\mathbb{L}}_{\dot2})\big|_{S^2}=0,\quad (\tpsi^{\mathbb{L}^\perp}_{\dot1} + \sigma_3\tpsi^{\mathbb{L}^\perp}_{\dot2})\big|_{S^2}=0,
\end{align}
and an additional one-loop determinant resulting from the localization along $S^2$:
\begin{equation}
Z_{\rm hyp}(\sigma,B, \mathbb{L}) = \prod_{w\in \mathbb{L}} (-1)^{\frac12(|w\cdot B| - w\cdot B)} \frac{\Gamma(\frac12 + iw\cdot\sigma + \frac{|w\cdot B|}{2})}{\Gamma(\frac12 - iw\cdot\sigma + \frac{|w\cdot B|}{2})}.
\end{equation}
The notation $q^{\mathbb L}_\pm = q^{\mathbb L}_1 \pm i q^{\mathbb L}_2$ refers to the components valued in $\mathbb{L}$, and $\tpsi^{\mathbb{L}^\perp}$ -- to the components valued in $\mathbb{L}^\perp$.

Thus in the end we obtain the final gluing formula:
\begin{equation}
Z_{S^3} = \sum_{B\in \Lambda_{\rm coch}} \frac1{|\cW(H_B)|} \int_{\mathfrak t} \dd^r\sigma\, \mu(\sigma, B, \mathbb{L}) Z^{(+)}_{HS^3}(\sigma, B, \mathbb{L}) Z^{(-)}_{HS^3}(\sigma, B, \mathbb{L}),
\end{equation}
where the gluing measure is
\begin{equation}\label{eq:gl-measure}
\mu(\sigma, B, \mathbb{L}) = Z_{\rm vm}(\sigma, B) Z_{\rm hyp}(\sigma,B, \mathbb{L}),
\end{equation}
and it explicitly depends on the choice of holomorphic Lagrangian $\mathbb{L}$. Here $Z^{(+)}_{HS^3}(\sigma, B, \mathbb{L})$ and $Z^{(-)}_{HS^3}(\sigma, B, \mathbb{L})$ mean the two hemisphere partition functions, with the boundary conditions described above that: (1) fix the magnetic flux $B$ through the boundary; (2) fix the scalar $\Phi_{\dot1\dot2}$ to be $\sigma/r$ along the boundary; (3) impose the $\mathbb{L}$-dependent $(2,2)$ boundary conditions on the hypermultiplets.

\paragraph{Dependence on $\mathbb{L}$.} We expect that the final answer, namely, the $S^3$ partition function and the collection of correlators, do not depend on the choice of $\mathbb{L}$, which we verify below and in the Appendix.

\section{Shift/Difference operators}\label{sec:shift-difference}
Now we need to determine the remaining ingredients: The hemisphere partition function or ``wavefunction'' $Z(\sigma, B)$ and the shift operators acting on it that represent the insertions of Coulomb branch operators on $HS^3$.

\subsection{$HS^3$ partition function}
The $HS^3$ partition function with the boundary conditions described above is computed like in \cite{Dedushenko:2017avn,Dedushenko:2018icp}. First we localize 3D $\cN=4$ vector multiplets using the Yang-Mills action as a $\cQ_C$-exact deformation. In the absence of monopole insertions, the latter sets all vectormultiplet fields to zero, except for the Cartan-valued constant vevs:
\begin{equation}
\Phi_{\dot1\dot2}=\frac{\sigma}{r},\quad D_{11} = D_{22} = -i\frac{\sigma}{r^2}.
\end{equation}
While on the full $S^3$ the answer is then written in terms of the integral over $\sigma\in\mathfrak{t}$, on $HS^3$ the boundary conditions fix $\Phi_{\dot1\dot2}=\sigma/r$, so no integration is left. The one-loop determinant for vectormultiplets on an empty $HS^3$ was computed in \cite{Dedushenko:2018icp}:
\begin{equation}
Z_{\rm vec}(\sigma, B) = \delta_{B,0} \prod_{\alpha\in\Delta} \frac{\sqrt{2\pi}}{\Gamma(1 - i\alpha\cdot \sigma)},
\end{equation}
where $\delta_{B,0}$ captures the fact that the hemisphere is empty, so there is no magnetic flux through the boundary. Next we localize the half-hypers. They are only coupled to the abelian background $\sigma\in \mathfrak{t}$, so it does not matter that they are half-hypers. Just like earlier, the half-hypers in $\mathbf{R}$ restricted to the maximal torus $T\subset G$ look like full hypers. Then $\mathbf{R}$ may be split into $T$-invariant subspaces: $\mathbb{L}\oplus \mathbb{L}^\perp$. The abelian hemisphere partition function was computed in \cite{Dedushenko:2017avn}, and the answer for empty $HS^3$ is:
\begin{equation}
Z_{\rm hyp}(\sigma, B, \mathbb{L}) = \delta_{B,0} \prod_{w\in \mathbb{L}} \frac{\Gamma(\frac12 - iw\cdot \sigma)}{\sqrt{2\pi}}.
\end{equation}
Thus the product $Z_{\rm vec}(\sigma, B) Z_{\rm hyp}(\sigma, B, \mathbb{L})$ computes the hemisphere partition function, with the $\cN=(2,2)$ Dirichlet boundary conditions on the vectormultiplet labeled by $(\sigma, B)$, and the $\cN=(2,2)$ boundary conditions on the half-hypers labeled by the Lagrangian splitting $\mathbf{R} = \mathbb{L} \oplus \mathbb{L}^\perp$. The Dirichlet boundary conditions on gauge fields reduce the gauge symmetry to a global symmetry $G$ along the boundary, which may be further broken to a subgroup by the vevs $(\sigma,B)$ and the splitting $\mathbf{R}=\mathbb{L}\oplus \mathbb{L}^\perp$.

The empty hemisphere is not enough, we also need an $HS^3$ partition function with a monopole operator of charge $b$ inserted at the pole. Computing it can be quite a grueling task, but luckily, all the hard work was done in \cite{Dedushenko:2017avn,Dedushenko:2018icp}. The localization (or BPS) equations consist of a group of equations on the vectormultiplet fields and those on the hypermultiplet fields. When we do the Coulomb branch localization (i.e., localization with respect to $\cQ_C$), the hypermultiplet BPS equations only have a zero solution, whether we insert a monopole or not. Interesting  things happen in the vectormultiplet sector.

The full set of localization equations on the vectormultiplet fileds can be found in \cite[eqn. 2.11]{Dedushenko:2018icp}. With the monopole singularity \eqref{eq:monopole_singularity} imposed, the solutions to these equations include non-trivial profiles for the gauge field and the vectormultiplet scalar $\Phi_{\dot1\dot1}$, while the scalar $\Phi_{\dot1\dot2}=\frac{\sigma}{r}$ remains constant (and hypermultiplets simply fluctuate on top of such a vectormultiplet background). One solution to these equations is the \emph{abelian solution} described in \cite[eqn. 2.12]{Dedushenko:2018icp}. Seeing the monopole charge (cocharacter) as an element of ${\rm Hom}(U(1), T)/\cW$, this cocharacter is represented by a Weyl orbit $\cW b$, where $b\in \mathfrak{t}$ is a dominant coweight determining a $T$-cocharacter $e^{i\varphi b} \in {\rm Hom}(U(1), T)$. As explained in detail in \cite{Dedushenko:2018icp}, after the localization the path integral reduces to a sum $\sum_{b'\in \cW b}$, where for each $b'$ we have a singularity \eqref{eq:monopole_singularity} and the corresponding solution \cite[eq. 2.12]{Dedushenko:2018icp}:
\begin{equation}\label{eq:abelian-solution}
\Phi_{\dot1\dot1}=\frac{ib'}{2r \sqrt{\cos^2\theta + \sin^2\theta \cos^2\varphi}},\quad A^\pm = \frac{b'}{2} \left(\frac{\sin\theta\cos\varphi}{\sqrt{1-\sin^2\theta \sin^2\varphi}}\pm 1\right)\dd\tau,
\end{equation}
where the sings $\pm$ correspond to the two patches covering the monopole configuration on $HS^3$. Notice that all these fields, in addition to
\begin{equation}
\Phi_{\dot1\dot2}=ir D_{11}= ir D_{22} = \frac{\sigma}{r},
\end{equation}
take values in the Cartan subalgebra $\mathfrak{t}\subset \mathfrak{g}$, so one simply has to compute the one-loop determinants on the abelian vectormultiplet background again. This is straightforward: The vectormultiplet contribution is copied from \cite{Dedushenko:2018icp}, and the half-hypers coupled to the abelian background forget that they are charged under $G$, acting as if they were full hypermultiplets in the representation $\mathbb{L} \oplus \mathbb{L}^\perp$ of the maximal torus $T\subset G$. Thus, we simply read off the answer from \cite{Dedushenko:2017avn,Dedushenko:2018icp}:
\begin{align}\label{eq:abelian-monopole-contribution}
Z_{HS^3}(b; \sigma, B, \mathbb{L}) &= \sum_{b'\in \cW b} [\text{phase}]\times \delta_{B,b'} \frac{\prod_{w\in\mathbb{L}} \frac1{\sqrt{2\pi}r^{|w\cdot B|/2}}\Gamma\left(\frac{1+|w\cdot B|}{2} - iw\cdot\sigma\right)}{\prod_{\alpha\in\Delta}\frac1{\sqrt{2\pi}r^{|\alpha\cdot B|/2}}\Gamma\left(1 + |\alpha\cdot B|/2 - i\alpha\cdot\sigma\right)}\cr
&\equiv \sum_{b'\in\cW b} Z_0(b'; \sigma, B, \mathbb{L}).
\end{align}
Here the denominator is the one-loop determinant of the vectormultipets, and the numerator -- of the hypermultiplets (which is the same as in \cite[eq. 2.15]{Dedushenko:2018icp}, with $\cR$ replaced by $\mathbb{L}$). The answer in \eqref{eq:abelian-monopole-contribution} includes an unknown phase factor, which could not be reliably determined from the one-loop determinant computation in \cite{Dedushenko:2017avn}. It was fixed there to be $[\text{phase}]=1$ from the consistency conditions, namely, the requirement that the monopole-antimonopole two-point function had the expected dependence on the monopole charge. However, that answer was only sensible for a fixed assignment of hypermultiplet charges. In a theory of full hypers, one may notice a symmetry: you can flip the sign of some hypermultiplet charges, or in the non-abelian case replace representation $\cR$ by its dual $\overline\cR$, without changing the Lagrangian! The answer \cite[eqn. 3.42]{Dedushenko:2017avn}, however, is not invariant under such a flip, possibly acquiring a minus sign. Thus, in order to account for this sign, we must allow in \eqref{eq:abelian-monopole-contribution} the possibility that $[\text{phase}]=\pm1$. Thus, borrowing the arguments from \cite{Dedushenko:2017avn}, the undetermined factor in \eqref{eq:abelian-monopole-contribution} is not really a phase, but just a sign. This sign should be adjusted in such a way that the Coulomb branch answers (the algebra $\cA_C$ and the correlators, when possible to compute) are invariant under replacing some of the matter representations by their conjugates. In our case, this sign will be adjusted so that the answer does not depend on $\mathbb{L}$.

\paragraph{Monopole bubbling.} The BPS equations \cite[eq. 2.11]{Dedushenko:2018icp} close to the monopole insertion (in fact, close to the $S^1_\varphi$ at $\theta=\frac{\pi}{2}$) include the Bogomoly equation $F = \star D u$, where $u= \Re (\tan\frac{\theta}{2} e^{i\varphi} \Phi_{\dot1\dot1})$. The nonabelian Bogomolny equations with the monopole singularity (say, inserted at the origin), besides the obvious abelian solution described above, admit bubbling solutions, as reviewed earlier. The monopole singularity $b$ gets screened to a smaller magnetic charge $v<b$ (where $v$ is a weight in the representation of the Langlands dual group ${}^L G$ of highest dominant weight $b$). Importantly, the moduli space of solutions to Bogomolny equations has a limit in which the ``screening radius'', behind which the monopole appears to carry a smaller magnetic flux $v<b$, goes to zero. In this limit, the screening takes place in the infinitesimal neighborhood of the monopole, outside of which the solution looks like the abelian solution described above.

Localization via the super Yang-Mills kinetic term requires the vectormultiplet BPS background to be abelian (due to the potential $-\tfrac14[\Phi^{\dot{a}}{}_{\dot b}, \Phi^{\dot c}{}_{\dot d}][\Phi^{\dot{b}}{}_{\dot a}, \Phi^{\dot d}{}_{\dot c}]$). Naively, this seems to forbid the bubbling solutions. However, the localization is done via the term $e^{-t S_{\rm YM}}$ in the limit $t\to\infty$, and at every finite $t$, we may mildly violate the BPS equations, so long as the functional $S_{\rm YM}$ deviates from $0$ by terms of order $o(t^{-1})$. The bubbling solutions are of this kind: they fail to be abelian in a tiny neighborhood of the monopole singularity, leading to an arbitrarily small $S_{\rm YM}$. Thus, they contribute in the localization procedure. In the limit $t\to\infty$, this gives additional localization loci, where the BPS solution looks like the abelian solution \eqref{eq:abelian-solution} of smaller charge $v<b$. The one-loop determinant on such a background is the same, possibly with additional effects captured by the ``bubbling factor'' $Z_0(b'\to v'; \sigma, B, \mathbb{L})$. Thus the full hemisphere partition function is:
\begin{equation}
Z_{HS^3}(b; \sigma, B, \mathbb{L}) = \sum_{b'\in \cW b} Z_0(b'; \sigma, B, \mathbb{L}) + \sum_{v<b}\sum_{\substack{b'\in\cW b\\ v'\in \cW v}} Z_0(b'\to v'; \sigma, B, \mathbb{L})Z_0(v'; \sigma, B, \mathbb{L}).
\end{equation}
These bubbling factors were subject of increased interest in the recent decade, as we reviewed in the Introduction. We do not attempt to extend our localization techniques to compute them. Instead, we follow the approach of \cite{Dedushenko:2018icp}, where it was argued that the bubbling factors can be bootstrapped from the algebraic consistency of $\cA_C$.

\subsection{Shift/Difference operators}
The insertions of Coulomb branch operators (along $HS^1 \subset HS^3$) can be represented by certain operators acting on the hemisphere partition function. Furthermore, since the half-circle $HS^1$ has two endpoints, those are the two ``entries'' through which a Coulomb branch operator $\cO$ can be added to $HS^1$, leading to the two ``versions'' of it: an operator $\cO_N$ representing the insertion at $\varphi=0$ and $\cO_S$ representing the insertion at $\varphi=\pi$:
\begin{figure}[h]
	\centering
	\includegraphics[width=0.3\textwidth]{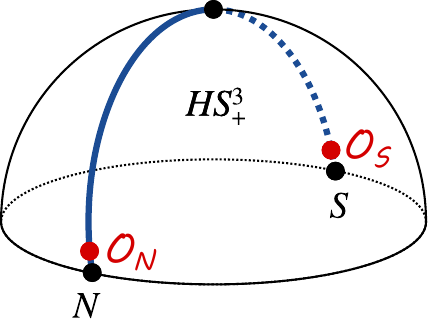}
	\caption{A twisted-translated Coulomb branch operator $\cO(\varphi)$ entering the hemisphere through the N endpoint of $HS^1$, that is $\cO(0)$, is represented by an operator $\cO_N$ acting on the hemisphere partition function, while $\cO(\pi)$ is represented by $\cO_S$.}\label{fig:hemi-insertions}
\end{figure}\\
Using these definitions, we can immediately derive $\Phi_N$ and $\Phi_R$ by relying on our boundary conditions:
\begin{align}
\Phi_N &= \Phi(0) = \frac12 (2\Phi_{\dot1\dot2} + \Phi_{\dot1\dot1} + \Phi_{\dot2\dot2})\big|_{\varphi=0} = \frac1{r} \left(\sigma + \frac{i}{2}B\right),\cr
\Phi_S &= \Phi(\pi) = \frac12 (2\Phi_{\dot1\dot2}-\Phi_{\dot1\dot1} - \Phi_{\dot2\dot2})\big|_{\varphi=\pi} = \frac1{r} \left(\sigma - \frac{i}{2}B\right).
\end{align}
As for the operators $M^b_N$ and $M^b_S$ representing monopoles, their form was derived in \cite{Dedushenko:2017avn,Dedushenko:2018icp} using several conditions: (1) for any $\cO$, the operators $\cO_N$ and $\cO_S$ commute (it does not make a difference whether you first insert an operator at point N or at S); (2) when acting on an empty hemisphere, $\cO^{(1)}_N \cO^{(2)}_N\dots \cO^{(n)}_N Z_0(0; \sigma, B, \mathbb{L}) = \cO^{(n)}_S \cO^{(n-1)}_S\dots \cO^{(1)}_S Z_0(0; \sigma, B, \mathbb{L})$, as moving the local operators from $\varphi=0$ to $\varphi=\pi$ reverses the order in which they act on the hemisphere; (3) naturally, we must have $Z_0(b; \sigma, B, \mathbb{L})= M_N^b Z_0(0; \sigma, B, \mathbb{L})$, i.e., $M_N^b$ adds a monopole operator that can be moved to the tip of the hemisphere, yielding the configuration that computes $Z_0(b; \sigma, B, \mathbb{L})$. These arguments were carefully spelled out in \cite[Section 5.1.1]{Dedushenko:2017avn}, and then generalized to the non-ableian case in \cite{Dedushenko:2018icp}. We see that, essentially, nothing changes in the case of theories with half-hypers, except the expression $Z_0(b; \sigma, B, \mathbb{L})$ now involves $\mathbb{L}$. Thus, we immediately conclude:
\begin{align}\label{eq:shift-op}
M_N^b &= [\text{sign}]\times \frac{\prod_{w\in\mathbb{L}}\frac{(-1)^{(w\cdot b)_+}}{r^{|w\cdot b|/2}}\left(\frac12 + ir w\cdot\Phi_N\right)_{(w\cdot b)_+}}{\prod_{\alpha\in\Delta} \frac{(-1)^{(\alpha\cdot b)_+}}{r^{|\alpha\cdot b|/2}}\left(ir \alpha\cdot\Phi_N\right)_{(\alpha\cdot b)_+}} e^{-b\cdot\left(\frac{i}{2}\partial_\sigma + \partial_B\right)},\cr
M_S^b &= [\text{sign}]\times \frac{\prod_{w\in\mathbb{L}}\frac{(-1)^{(-w\cdot b)_+}}{r^{|w\cdot b|/2}}\left(\frac12 + ir w\cdot\Phi_S\right)_{(-w\cdot b)_+}}{\prod_{\alpha\in\Delta} \frac{(-1)^{(-\alpha\cdot b)_+}}{r^{|\alpha\cdot b|/2}}\left(ir \alpha\cdot\Phi_S\right)_{(-\alpha\cdot b)_+}} e^{b\cdot\left(\frac{i}{2}\partial_\sigma - \partial_B\right)},
\end{align}
where $[\text{sign}]$ represents the sign ambiguity, the leftover of the undetermined phase in \eqref{eq:abelian-monopole-contribution} discussed after \eqref{eq:abelian-monopole-contribution}. We also use notations: $(x)_+ = (x + |x|)/2$, and $(x)_n = \Gamma(x+n)/\Gamma(x) = x(x+1)\dots(x+n-1)$. The operator $e^{-b\cdot \partial_B}$ is a shift operator that, because of
\begin{equation}
e^{-b\cdot \partial_B} \delta_{B,0} = \delta_{B-b,0} = \delta_{B,b},
\end{equation}
reflects the fact that both $M^b_N$ and $M^b_S$ add a magnetic flux $b$ to the hemisphere. The operator $e^{\pm \frac{i}{2} b\cdot\partial_\sigma}$ shifts $\sigma$ by $\pm \frac{i}{2} b$ in the argument of the hemisphere wavefunction. Collecting the total power of $r$, we get the familiar formula for monopole scaling dimension:
\begin{equation}
\Delta_b = \frac12 \left(\sum_{w\in\mathbf{L}} |w\cdot b| - \sum_{\alpha\in\Delta}|\alpha\cdot b|\right) = \frac14 \sum_{w\in\mathbf{R}} |w\cdot b| - \frac12 \sum_{\alpha\in\Delta}|\alpha\cdot b|,
\end{equation}
where the second expression is written in terms of the full pseudoreal representation $\mathbf{R}$ to emphasize that the answer is independent of the splitting $\mathbf{R} = \mathbb{L} \oplus \mathbb{L}^\perp$. Here we used the fact that in every pair of weights $(w, -w)$, one must belong to $\mathbb{L}$ and the other -- to $\mathbb{L}^\perp$, as the symplectic pairing of such weights is non-zero, so they cannot belong to the same Lagrangian subspace.

\paragraph{The sign.} To fix the unknown $[\text{sign}]$ in \eqref{eq:shift-op}, we must require that the two-point function represented by $M_N^{-b} M_N^b Z_0(0; \sigma, B, \mathbb{L})=M_S^{b} M_S^{-b} Z_0(0; \sigma, B, \mathbb{L})$ is invariant under the ``flipping'' of charges, which also means that it is independent of $\mathbb{L}$: choosing $\mathbb{L}$ amounts to picking, for each pair of weights $(w, -w)$, whether we add $w$ or $-w$ to $\mathbb{L}$ (i.e., in the abelianized description, choosing the sign of $T$-charge of each abelian hypermultiplet). One choice that works for us is:
\begin{equation}
[\text{sign}] = (-1)^{\frac12 \left(\sum_{w\in\mathbb{L}} w\cdot b\right)_+},
\end{equation}
which is indeed just a sign if the theory is non-anomalous.

General dressed monopoles, including the bubbling terms, are represented as:
\begin{equation}\label{eq:general-shift-op}
[P(\Phi)\cM^b] = \sum_{{\rm w}\in \cW} P(\Phi_N^{\rm w}) M_N^{{\rm w}\cdot b} + \sum_{{\rm w}\in\cW}\sum_{\substack{v<b\\ v'\in \cW v}} P(\Phi_N^{\rm w})Z_0(b^{\rm w}\to v'; \sigma, B, \mathbb{L})M_N^{{\rm w}\cdot v'}.
\end{equation}
Here both $X^{\rm w}$ and ${\rm w}\cdot X$ mean the same thing: the action of the Weyl group element ${\rm w}$ on $X \in \mathfrak{t}$. The formula \eqref{eq:general-shift-op} is written in the ``N'' representation, but it equally well may be written in the ``S'' representation (and the precise form of the bubbling factors $Z_0$ may be different for the two representations).

\subsection{Algebraic consistency, or polynomiality and mixing}\label{sec:polynomiality}
One key idea in the nonabelian story of \cite{Dedushenko:2018icp} was the \emph{hypothesis of polynomiality}, leading to the \emph{conjecture} of polynomiality verified there for theories with gauge groups of rank $\leq 2$. As explained in the Introduction, the hypothesis of polynomiality states that the Coulomb branch algebra $\cA_C$ only has polynomial relations. 
This prohibits any denominators from appearing, which is quite non-trivial, since the shift operators \eqref{eq:shift-op} have denominators, and imposes strong constraints on the form of bubbling coefficients.

The conjecture of polynomiality states that this condition fully determines the form of bubbling coefficients up to the choice of basis in the space of local operators (the mixing ambiguity). In a flat space CFT, operators of the same scaling dimension can mix. In a theory with dimensionful scale, the operators of different scaling dimension can mix too, the difference being compensated by the power of a dimensionful parameter (the UV scale, the IR scale, or some coupling). In our case the relevant parameter is the radius $r$ of the sphere. The operators may mix with those of lower dimension, the difference compensated by $r$:
\begin{equation}
\cO = \cO_\Delta + \sum_{\Delta'\leq\Delta} \frac1{r^{\Delta-\Delta'}} \cO_{\Delta'},
\end{equation}
where $\cO_{\Delta}$ has dimension $\Delta$. The ``mixing'' here is not a physical phenomenon, but rather an artifact of description. For a randomly constructed operator, the mixing is most certainly present and must be resolved via the Gram-Schmidt procedure, as was done, e.g., in \cite{Dedushenko:2016jxl}.

While the basis of operators of definite dimension is often preferred, it may certainly happen that in the process of solving the theory, other bases also play role. This is the case for us: The basis that simplifies the bubbling coefficients helps a lot. Roughly, we constrain the bubbling coefficients using the polynomiality hypothesis and up to the freedom of operator mixing. The remarkable feature is that this fully determines the bubbling coefficients, hence solves the theory. After that, if needed, we can, of course, diagonalize the two-point functions to recover the CFT data. Let us describe this in slightly more detail.

\paragraph{Theories with minuscule monopoles.} The monopole charges are Weyl orbits in $\Lambda^\vee$, the weight lattice of the dual group ${}^L G$. Pick the dominant weight $b\in\Lambda^\vee$, consider the ${}^L G$-module $R_b$ of highest weight $b$. The monopole of charge $b$ can only bubble to monopoles whose charges also belong to $R_b$ \cite{Kapustin:2006pk,Gomis:2011pf}, except those in the Weyl orbit $\cW b$. When the full set of weights in $R_b$ coincides with the Weyl orbit $\cW b$---such representations are called \emph{minuscule}---there is no bubbling. Such monopoles are represented by $\sum_{\rm w} M^{{\rm w}\cdot b}$. Whether a theory admits minuscule monopoles or not is a condition on the global form of $G$. Oftentimes, the minuscule monopoles actually generate the whole algebra $\cA_C$, and we are done (even though the higher-charge monopoles bubble). The basic example is $G=U(N)$, as ${}^L G = U(N)$, and its $N$-dimensional representation is minuscule. Coulomb branches of theories with $G=U(N)$ (and unitary quivers) were the first well-understood examples \cite{Bullimore:2015lsa}. Another basic example is $G=SO(3)$: for ${}^L G=SU(2)$, the two-dimensional representation is minuscule, as its weights form a single Weyl orbit. This monopole does not bubble, and furthermore, $M^1 + M^{-1}$ and $\Phi(M^1-M^{-1})$ generate the whole algebra, so the Coulomb branch sector is again solved without ever analyzing the bubbling. But in any case, neither $G=U(N)$ nor $G=SO(3)$ admit any pseudoreal representations, so such theories only have full hypers, and all the answers can be found in \cite{Dedushenko:2018icp}.

\paragraph{True bubbling.} Now we will study theories that have half-hypermultiplets and whose minimal monopoles bubble. The main example is $G=SU(2)$, which has both properties. Indeed, ${}^L G= SO(3)$, and its minimal representation is three-dimensional, whose weights $\{-2, 0, 2\}$ split into two Weyl orbits, $\{-2, 2\}$ and $\{0\}$. The monopole characterized by the orbit $\{-2, 2\}$ may bubble into the charge-zero monopole, i.e., exhibit a complete screening. This is reflected in the fact that the correct abelianized monopole is:
\begin{align}
\widetilde{M}^2 = M^2 + Z(\Phi),
\end{align}
where $Z(\Phi) = Z_0(2 \to 0, \sigma, B, \mathbb{L})$ is the bubbling term which is a function of $\Phi$ only.

Let us repeat the argument of \cite{Dedushenko:2018icp}. Shifting $\widetilde{M}^2$ (or rather its Weyl-average $\widetilde{M}^2 + \widetilde{M}^{-2}$) by the lower-dimension operators of monopole charge zero, such as (Weyl-invariant) polynomials $P(\Phi)$, we change the form of $Z(\Phi)$, i.e.\ the functional form of $Z(\Phi)$ depends on the choice of basis of operators. Consider the bubbled minimal monopole and its dressed version:
\begin{align}
\cM^2 &= M^2 + M^{-2} + Z(\Phi) + Z(-\Phi),\cr
[\Phi\cM^2] &= \Phi(M^2 - M^{-2}) + \Phi(Z(\Phi) - Z(-\Phi)).
\end{align}
A monopole of the same charge dressed by a general polynomial $F(\Phi)= f(\Phi^2) + g(\Phi^2)\Phi$ can be written as $[F(\Phi)\cM^2] = f(\Phi^2) \cdot \cM^2 + g(\Phi^2) \cdot [\Phi \cM^2]$, and monopoles of higher charge can be obtained by taking products. Thus $\cM^2$, $[\Phi \cM^2]$ and $\Phi^2$ generate the whole $\cA_C$. However, $\cM^2$ and $[\Phi\cM^2]$ include the unknown function $Z(\Phi)$. In order to resolve this issue, we pick a different set of generators: $\Phi^2$, $X$ and $Y$, where
\begin{equation}
X = \cM^2 + P_1(\Phi^2),\quad Y = [\Phi\cM^2] + P_2(\Phi^2),
\end{equation}
with some unknown polynomials $P_1$ and $P_2$ that account for the mixing ambiguity. We are about to argue that, while the form of $Z(\Phi)$, $P_1(\Phi)$, and $P_2(\Phi)$ remain undetermined, there exist unique choices of $P_1$ and $P_2$ such that $X$ and $Y$ admit explicit expressions. Thus we will use $\Phi^2$, $X$, $Y$ as generators, yielding an explicit description of $\cA_C$.

The monopole shift operators (written in the N representation) are:
\begin{align}\label{eq:su2_M2}
M^2 &= (-1)^{\left(\sum_{w\in\mathbb{L}} w\right)_+}\frac{\prod_{w\in\mathbb{L}} \frac{(-1)^{(2w)_+}}{r^{|w|}} \left(\frac12 + irw\Phi\right)_{(2w)_+}}{\frac1{r^2} (ir\Phi+1)ir\Phi} e^{-2\partial_B - i\partial_\sigma}, \\
M^{-2} &= (-1)^{\left(-\sum_{w\in\mathbb{L}} w\right)_+}\frac{\prod_{w\in\mathbb{L}} \frac{(-1)^{(-2w)_+}}{r^{|w|}} \left(\frac12 + irw\Phi\right)_{(-2w)_+}}{\frac1{r^2} (ir\Phi-1)ir\Phi} e^{2\partial_B + i\partial_\sigma}.
\end{align}
Compute the following commutators of shift operators:
\begin{align}
&X\Phi^2 - \Phi^2 X = 
- \frac{4i}{r} Y - \frac{4}{r^2} X\cr &+ \frac4{r^2}(Z(\Phi) + Z(-\Phi) + P_1(\Phi^2)) + \frac{4i}{r} \left(\Phi(Z(\Phi) - Z(-\Phi)) + P_2(\Phi^2)\right),\\
&Y\Phi^2 - \Phi^2 Y = - \frac{4i}{r} \Phi^2 X - \frac{4}{r^2} Y\cr &+ \frac4{r^2}\left(\Phi(Z(\Phi) - Z(-\Phi)) + P_2(\Phi^2)\right) + \frac{4i}{r}\Phi^2 \left(Z(\Phi) + Z(-\Phi) + P_1(\Phi^2)\right).
\end{align}
Polynomiality hypothesis tells us that the right hand side is polynomial. In particular:
\begin{align}
&\frac1r(Z(\Phi) + Z(-\Phi) + P_1(\Phi^2)) + i\left(\Phi(Z(\Phi) - Z(-\Phi)) + P_2(\Phi^2)\right) = i A_0(\Phi^2),\\
&\frac1r\left(\Phi(Z(\Phi) - Z(-\Phi)) + P_2(\Phi^2)\right) + i\Phi^2 \left(Z(\Phi) + Z(-\Phi) + P_1(\Phi^2)\right) = i A_1(\Phi^2)
\end{align}
are polynomials in $\Phi^2$. From the first equation, we fix $P_2(\Phi^2)$ such that $A_0(\Phi^2)=0$:
\begin{equation}
P_2(\Phi^2) + \Phi(Z(\Phi) - Z(-\Phi)) = \frac{i}{r}\left(Z(\Phi) + Z(-\Phi) + P_1(\Phi^2)\right).
\end{equation}
Then the second equation becomes:
\begin{align}
\left(\frac{1}{r^2} + \Phi^2\right)\left(Z(\Phi) + Z(-\Phi) + P_1(\Phi^2)\right) = A_1(\Phi^2)
\end{align}
Since we can always write $A_1(\Phi^2)=c + (r^{-2} + \Phi^2)B(\Phi^2)$, we can clearly choose $P_1(\Phi^2)$ in such a way that $A_1(\Phi)=c$ is a constant, so:
\begin{align}
Z(\Phi) + Z(-\Phi) + P_1(\Phi^2) &= \frac{c}{\Phi^2 + \frac{1}{r^2}},\cr
P_2(\Phi^2) + \Phi(Z(\Phi) - Z(-\Phi)) &= \frac{i}{r}\frac{c}{\Phi^2 + \frac{1}{r^2}}.
\end{align}
Thus we fix the expression for $X$ and $Y$ uniquely, up to a single constant $c$:
\begin{align}
X = M^2 + M^{-2} + \frac{c}{\Phi^2 + \frac{1}{r^2}},\cr
Y = \Phi(M^2 - M^{-2}) + \frac{i}{r}\frac{c}{\Phi^2 + \frac{1}{r^2}}.
\end{align}
Is the algebra $\cA_C$ defined for every $c$? No, as it turns out, the polynomiality hypothesis further constrains $c$. For that, use the shift operators \eqref{eq:su2_M2} to compute the following:
\begin{align}\label{eq:constraint-c}
X Y - YX + \frac{2i}{r}XX = (-1)^{|\sum_{w\in\mathbb{L}}w|} \frac{2\prod_{w\in\mathbb{L}} \frac{(-1)^{2|w|}}{r^{2|w|}}(\frac12 + irw\Phi)_{(-2w)_+}(\frac12 -2w + irw\Phi)_{(2w)_+}}{\Phi (\Phi+\frac{i}{r})^2}\cr - (-1)^{|\sum_{w\in\mathbb{L}}w|} \frac{2\prod_{w\in\mathbb{L}} \frac{(-1)^{2|w|}}{r^{2|w|}}(\frac12 + irw\Phi)_{(2w)_+}(\frac12 + 2w + irw\Phi)_{(-2w)_+}}{\Phi (\Phi-\frac{i}{r})^2} + \frac{2i}{r} \frac{c^2}{\left(\Phi^2 + \frac1{r^2}\right)^2}.
\end{align}
The right-hand side should be a Weyl-invariant polynomial in $\Phi$. In particular, all the poles must cancel. For $G=SU(2)$, the weights are valued in $\frac12 \Z$: Those in $\Z$ belong to the integer-spin representations that are real, while those in $\Z + \frac12$ belong to pseudoreal representations (which must be present if we want to have half-hypers). Let us introduce:
\begin{equation}
\sum_{w\in \mathbb{L}} |w| = A,
\end{equation}
Notice that $2A\mod 2$ is precisely the $\Z_2$ anomaly discussed in Section \ref{sec:anomaly}. With this notation, $\prod_w \frac{(-1)^{2|w|}}{r^{2|w|}}=\frac{(-1)^{2A}}{r^{2A}}$. Now note the following equality:
\begin{equation}
\left(\frac12 - 2w + irw\Phi\right)_{(2w)_+} = (-1)^{(2w)_+} \left(\frac12 - irw\Phi\right)_{(2w)_+}
\end{equation}
which is verified using the definition $(x)_n=x(x+1)\dots(x+n-1)$. We apply this relation to \eqref{eq:constraint-c}, use $(-1)^{(2w)_+ + (-2w)_+}=(-1)^{2A}$, and express the right hand side of \eqref{eq:constraint-c} as:
\begin{align}\label{eq:poles-cancel}
\frac{2}{r^{2A}\Phi}\Big[\chi(\Phi) - (-1)^{2A} \chi(-\Phi)\Big] + \frac{2i}{r} \frac{c^2}{\left(\Phi^2 + \frac1{r^2}\right)^2}.
\end{align}
where we introduced:
\begin{equation}
\chi(\Phi) = \mathfrak{s} \times \frac{\prod_{w\in\mathbb{L}} (\frac12 - irw\Phi)_{(2w)_+}(\frac12 + 2w - irw\Phi)_{(-2w)_+}}{(\Phi+\frac{i}{r})^2},\quad \mathfrak{s}=(-1)^{|\sum_{w\in\mathbb{L}}w|}.
\end{equation}
This shows that when $(-1)^{2A}=1$ (i.e.\ the $\Z_2$ anomaly is zero), the pole at $\Phi=0$ cancels in \eqref{eq:poles-cancel}. Curiously, for $(-1)^{2A}=-1$, this pole cannot vanish and the polynomiality fails. This is what goes wrong with our formalism when the $\Z_2$ anomaly is present:
\begin{equation*}
\boxed{\text{In anomalous theory, the hypothesis of polynomiality fails.}}
\end{equation*}
The expression \eqref{eq:poles-cancel} also has apparent poles at $\Phi = \pm \frac{i}{r}$, which both must cancel. To extract the pole at $\Phi=-\frac{i}{r}$, we set $\Phi =-\frac{i}{r} + x$ and only keep the terms that contribute to the pole:
\begin{equation}
\frac{2\mathfrak{s}}{r^{2A}} \frac{\prod_{w\in\mathbb{L}} \left(\frac12 - w - irw x\right)_{(2w)_+}\left(\frac12 +w-irw x\right)_{(-2w)_+}}{\left(-\frac{i}{r} +x\right)x^2} + \frac{2i}{r} \frac{c^2}{(\frac{2i}{r}-x)^2x^2}.
\end{equation}
Notice that $(\frac12-w-irwx)_{2w}=(\frac12-w-irwx)(\frac32-irwx)\dots(w-\frac12-irwx)$, so collecting the first with the last factors, the next-to-first with the next-to-last, and so on, gives us $\prod_i\left[-\left(\frac12 + i -w\right)^2-r^2w^2x^2 \right]$ that only depends on $x^2$. Thus in the numerator, we can simply set $x=0$ (all other terms are $O(x^2) $ and do not contribute to pole). Expanding the rest in $\frac1x$, we collect the following polar terms:
\begin{align}
\left(\frac{r}{2i x^2} - \frac{r^2}{2 x}\right)\left(c^2 - \frac{4\mathfrak{s}}{r^{2A}}\prod_{w\in\mathbb{L}}\left(\frac12-w\right)_{(2w)_+}\left(\frac12+w\right)_{(-2w)_+}\right),
\end{align}
clearly implying that
\begin{equation}
c^2 = \frac{4\mathfrak{s}}{r^{2A}}\prod_{w\in\mathbb{L}}\left(\frac12-w\right)_{(2w)_+}\left(\frac12+w\right)_{(-2w)_+} \equiv \frac{4\mathfrak{s}}{r^{2A}}\prod_{w\in\mathbb{L}}\left(\frac12-|w|\right)_{2|w|}.
\end{equation}
Analyzing the pole at $\Phi=\frac{i}{r}$ gives the same result. If a theory has at least one half-integral weight, then the above expression clearly vanishes:
\begin{equation}
\mathbb{L} \cap \left(\Z+\frac12\right) \neq \emptyset \Rightarrow c=0.
\end{equation}
In particular, a theory with genuine half-hypermultiplets must contain pseudoreal representations, i.e.\ those of half-integer $SU(2)$ spin. For such theories, $c=0$. When all the weights are integers, we get a non-trivial $c$ above, and the sign of $c$ can be fixed with the trick used in \cite{Dedushenko:2018icp} (start with the $U(2)$ gauge group, construct the monopole and gauge $U(1)_{\rm top}$).

More generally, assume that the matter, besides being charged under $SU(2)$, is coupled to some other group $G'$, global or gauge. Consider a representation $\mathbf{s}\otimes R'$ of $SU(2)\times G'$, where $\mathbf{s}$ is an $SU(2)$ irrep of spin $s$. When $s\in \Z +\frac12$, it is pseudoreal and we can describe this as follows: There are $N$ half-hypers in $\mathbf{s}$, with global symmetry $SO(N)$ and $G'\subset SO(N)$. Then choose a Lagrangian in $\mathbf{s}\otimes R'$ simply as $\mathbb{L}\otimes R'$, where $\mathbb{L}\subset \mathbf{s}$ is Lagrangian, and any weight $(w,w')\in \mathbb{L}\otimes R'$ comes together with $(w, -w')$. The other case is $s\in\Z$, so $\mathbf{s}$ is real and $R'$ must be pseudoreal. In this case choose $\mathbf{s}\otimes \mathbb{L}'$ as a Lagrangian, where $\mathbb{L}'\in R'$ is Lagrangian. In either case, introduce a scalar $\Phi'$ (dynamical of background) that couples to $w'$. Then the expressions for $M^2$ and $M^{-2}$ in \eqref{eq:su2_M2} involve products over $(w,w')$, an we replace each $w\Phi$ by $w\Phi + w'\Phi'$. Pretty much the same analysis goes through, leading to the necessity of pole cancellation in:
\begin{align}
\frac{2\mathfrak{s}\prod_{(w,w')} \frac{(-1)^{2|w|}}{r^{2|w|}}(\frac12 + irw\Phi + irw'\Phi')_{(-2w)_+}(\frac12 -2w + irw\Phi + irw'\Phi')_{(2w)_+}}{\Phi (\Phi+\frac{i}{r})^2}\cr
- \frac{2\mathfrak{s}\prod_{(w,w')} \frac{(-1)^{2|w|}}{r^{2|w|}}(\frac12 + irw\Phi + irw'\Phi')_{(2w)_+}(\frac12 + 2w + irw\Phi + irw'\Phi')_{(-2w)_+}}{\Phi (\Phi-\frac{i}{r})^2} + \frac{2i}{r} \frac{c^2}{\left(\Phi^2 + \frac1{r^2}\right)^2}.
\end{align}
Since $(w,w')$ is paired either with $(w, -w')$ or $(-w, w')$, we again show that the $\Phi=0$ pole cancels, while the cancellation of $\Phi=\pm \frac{i}{r}$ poles leads to the similar result, if the anomaly cancellation condition $(-1)^{2A}=1$ holds:
\begin{equation}
c^2 = \frac{4(-1)^{|\sum_{w\in\mathbb{L}}w|}}{r^{2A}} \prod_{(w,w')} \left(\frac12 - |w| + irw' \Phi'\right)_{2|w|}.
\end{equation}
One can also easily see (since every $(w,w')$ is paired with $(-w,w')$ or $(w,-w')$) that this expression is a total square, so that $c$ is a polynomial in $\Phi'$.

\section{Examples}\label{sec:examples}
Now we have all the tools to handle a few examples of theories with half-hypermultiplets, in which we are able to fully solve the Coulomb branch sector.
\subsection{General $SU(2)$ with half-hypers}
In a theory with gauge group $SU(2)$ and at least one half-integer spin representation (which includes all examples with the half-hypers), as we have seen above, $c=0$. All the Coulomb branch operators are built from $\Phi^2$ (as $\Phi\mapsto -\Phi$ under the Weyl group), where we take $\Phi\equiv \Phi_N = \frac1r (\sigma + \frac{i}{2}B)$, and the two minimal Weyl-invariant monopoles:
\begin{align}
X &= M^2 + M^{-2},\cr
Y &= \Phi(M^2 - M^{-2}).
\end{align}
We can check that for $b_1, b_2>0$, $M^{b_1} M^{b_2} = M^{b_1+ b_2}$ and $M^{-b_1}M^{-b_2}=M^{-b_1-b_2}$, thus $M^{2n}= (M^2)^n$ and $M^{-2n}=(M^{-2})^n$. This proves that $X$, $Y$ and $\Phi^2$ are enough to generate everything: $X^n$ will contain $M^{2n} + M^{-2n}$ plus lower-dimensional terms, while $Y X^{n-1}$ will contain $\Phi(M^{2n} - M^{-2n})$ plus lower-dimensional terms. From these, we construct the basis of the algebra of shift operators (viewed as a complex vector space):
\begin{equation}
\text{Basis:}\quad \Phi^{2m}X^n,\ m\geq0,\ n\geq0,\quad \text{and}\quad \Phi^{2m}Y X^n,\ m\geq0,\ n\geq0.
\end{equation}
Let us determine the relations that allow to write any element as a linear combination of these. Two relations are commutators that are independent of the precise matter content:
\begin{align}\label{eq:XPh-comm}
X\Phi^2 - \Phi^2 X &=  -\frac{4i}{r}Y - \frac{4}{r^2} X,\\
\label{eq:YPh-comm}Y\Phi^2 - \Phi^2 Y &=  -\frac{4i}{r} \Phi^2 X - \frac{4}{r^2} Y.
\end{align}
One relation is a commutator that depends on the matter content:
\begin{align}
\label{eq:XY-comm}XY - YX &= -\frac{2i}{r} X^2 + \frac{1}{r} p(\Phi),\\ 
p(\Phi) &= 4i(M^2 M^{-2} + M^{-2} M^2) + 2r\Phi (M^{-2}M^2 - M^2 M^{-2}),
\end{align}
where the matter-dependent piece $p(\Phi)$ evaluates to:
\begin{align}
p(\Phi)=\frac{2 \mathfrak{s}\prod_{w\in\mathbb{L}}(-1)^{(2w)_+}}{r^{2A-3}} \times \frac1\Phi \left[\frac{\prod_{w\in\mathbb{L}}\left(\frac12 +ir|w|\Phi\right)_{|2w|}}{(1+ir\Phi)^2}-\frac{\prod_{w\in\mathbb{L}}\left(\frac12 -ir|w|\Phi\right)_{|2w|}}{(1-ir\Phi)^2}\right],
\end{align}
which is clearly a polynomial in $\Phi^2$: the poles at $\Phi=\pm\frac{i}{r}$ cancel because a non-anomalous theory with half-hypers has at least two half-integral weights $w\in\mathbb{L}$; and the pole at $\Phi=0$ cancels due to the symmetry under $\Phi\mapsto -\Phi$, which also shows that it is a function of $\Phi^2$. The final and most important relation allows to decrease the power of $Y$ in any expression:
\begin{align}\label{eq:Y2toY}
Y^2 &= \Phi^2 X^2 - \frac{2i}{r} YX + \mu(\Phi)\\
\mu(\Phi) &= -2\Phi^2(M^2 M^{-2} + M^{-2}M^2) -\frac{4i}{r} \Phi(M^{-2}M^2 - M^2 M^{-2}),
\end{align}
where again the matter-dependent part $\mu(\Phi)$ is a certain polynomial in $\Phi^2$:
\begin{equation}
\mu(\Phi) = \frac{2 \mathfrak{s}\prod_{w\in\mathbb{L}}(-1)^{(2w)_+}}{r^{2A-2}}  \left[\frac{\prod_{w\in\mathbb{L}}\left(\frac12 +ir|w|\Phi\right)_{|2w|}}{(1+ir\Phi)^2}+\frac{\prod_{w\in\mathbb{L}}\left(\frac12 -ir|w|\Phi\right)_{|2w|}}{(1-ir\Phi)^2}\right].
\end{equation}
It is instructive to take the commutative limit $r\to\infty$, in which $p(\Phi)$ vanishes and $\mu(\Phi)$ becomes (using that $A\in\Z$):
\begin{equation}
\lim_{r\to\infty} \mu(\Phi) =(-\Phi^2)^{A-1} 4\mathfrak{s}\prod_{w\in\mathbb{L}}(-1)^{(2w)_+}|w|^{2|w|} = -\alpha (\Phi^2)^{A-1}.
\end{equation}
Thus we obtain commuting variables $X, Y, Z=\Phi^2$ subject to the relation:
\begin{equation}
Y^2 = Z X^2 - \alpha Z^{A-1}.
\end{equation}
We can always set $\alpha=1$ by rescaling $X$ and $Y$, obtaining the usual form of the $D_{A}$ singularity, where, recall, $A=\sum_{w\in\mathbb{L}}|w|$. A special case of this result was obtained in \cite{Dedushenko:2018icp} for a theory with $N_f$ full fundamental hypers and $N_a$ adjoint ones, where it was found that $A=N_f + 2N_a$, indeed matching our answer.

\subsection{Theories with the $D_2$ Coulomb branch}
Whenever $\sum_{w\in\mathbb{L}}|w|=2$, we get the Coulomb branch that after rescaling the variables looks like $y^2 = zx^2 - z$, which, formally, is a $D_2$ singularity. The latter is not on the list of singularities: it is not even singular at the origin. Instead, it has two isolated $A_1$ singularities at $x=\pm1$, $y=z=0$, reflecting that at the level of Dynkin diagrams, $D_2 = A_1 \sqcup A_1$.

There are three non-anomalous $SU(2)$ gauge theories with $A=2$: (1) theory with a single spin-$\frac32$ half-hyper; (2) theory with two full fundamental (spin-$\frac12$) hypers mentioned in \cite{Assel:2018exy,Dedushenko:2018icp,Giacomelli:2024laq}; (3) theory with a single adjoint full hyper. All of these are ``bad'' theories, have isomorphic Coulomb branches, yet different Higgs branches. Only the first theory contains a genuine half-hyper, while the other two consist of full hypers and are amenable to the older methods. Note that the last theory actually has $\cN=8$ (more on that below).

\subsubsection{$SU(2)$ with spin-$\frac32$ half-hyper}
First consider an $SU(2)$ gauge theory with one half-hypermultiplet in a spin-$\frac32$ (i.e., four-dimensional) irrep. This is the simplest example with a half-hyper (since a single spin-$\frac12$ is anomalous). There are four ways to pick the weights spanning the Lagrangian subspace:
\begin{align}
\mathbb{L}_{++} &= \{\frac12, \frac32\},\\
\mathbb{L}_{+-} &= \{\frac12, -\frac32\},\\
\mathbb{L}_{-+} &= \{-\frac12, \frac32\},\\
\mathbb{L}_{--} &= \{-\frac12, -\frac32\}.
\end{align}
All four cases, referred to as $++$, $+-$, $-+$ and $--$, lead to the same answers. We find:
\begin{align}
++:\ &M^2 = \frac{(1 + 3ir\Phi)(3 + 3ir\Phi)(5 + 3ir\Phi)}{16ir\Phi} e^{-\frac{2i}{r} \partial_\Phi},\quad M^{-2} = \frac{1}{(ir\Phi-1)ir\Phi} e^{\frac{2i}{r}\partial_\Phi},\\
+-:\ &M^2 = -\frac{1}{2ir\Phi} e^{-\frac{2i}{r}\partial_\Phi},\quad M^{-2} = -\frac{3(1 - 3ir\Phi)(5 - 3ir\Phi)}{8ir\Phi} e^{\frac{2i}{r}\partial_\Phi},\\
-+:\ &M^2 = \frac{3(1 + 3ir\Phi)(5 + 3ir\Phi)}{8ir\Phi} e^{-\frac{2i}{r}\partial_\Phi},\quad M^{-2} = \frac{1}{2ir\Phi} e^{\frac{2i}{r}\partial_\Phi},\\
--:\ &M^2 = \frac{1}{(ir\Phi+1)ir\Phi} e^{-\frac{2i}{r} \partial_\Phi},\quad M^{-2} = -\frac{(1 - 3ir\Phi)(3 - 3ir\Phi)(5 - 3ir\Phi)}{16ir\Phi} e^{\frac{2i}{r}\partial_\Phi}.
\end{align}
The three commutation relations specialize to:
\begin{align}\label{eq:XPh-comm32}
X\Phi^2 - \Phi^2 X &=  -\frac{4i}{r}Y - \frac{4}{r^2} X,\\
\label{eq:YPh-comm32}Y\Phi^2 - \Phi^2 Y &=  -\frac{4i}{r} \Phi^2 X - \frac{4}{r^2} Y,\\
\label{eq:XY-comm32}XY - YX &= -\frac{2i}{r} X^2 + \frac{27i}{2r}.
\end{align}
And the $Y^2$ relation reads:
\begin{equation}\label{eq:Y2toY32}
Y^2 = \Phi^2 X^2 - \frac{2i}{r} YX - \frac{27}{4} \Phi^2 + \frac{15}{4r^2},
\end{equation}
completing the description of $\cA_C$ as a free associative algebra generated by $X, Y, Z=\Phi^2$, modulo the relations \eqref{eq:Y2toY32}-\eqref{eq:XY-comm32}:
\begin{equation}
\cA_C = \C\langle X,Y,\Phi^2\rangle/(\eqref{eq:Y2toY32}, \eqref{eq:XPh-comm32}, \eqref{eq:YPh-comm32}, \eqref{eq:XY-comm32}).
\end{equation}
While $Y$ has dimension $1$ and $\Phi^2$ is a dimension $2$ operator, $X$ has dimension $0$, indicative of the ``bad'' theory. In the commutative limit $r\to\infty$, we recover the Coulomb branch as a variety determined by:
\begin{equation}
Y^2 = Z X^2 - \frac{27}{4} Z.
\end{equation}
This surface, though formally a $D_2$ singularity, is smooth at the origin $X=Y=Z=0$ (indeed, there is no such thing as $D_2$ singularity). It has two isolated $A_1$ singularities (and no other singular points) at:
\begin{equation}
X = \pm \frac{3\sqrt{3}}{2},\ Y=Z=0.
\end{equation}

These singularities must support some SCFT, however, notably, they do \emph{not} represent intersection with the Higgs branch, for in this example, there is no Higgs branch, since the hyper-K\"{a}hler quotient is trivial:
\begin{equation}
\cM_C=\C^4///SU(2)=0.
\end{equation}
The SCFT at either singularity must have the Coulomb branch isomorphic to the $A_1$ singularity and no Higgs branch. It could be a $\Z_2$ gauging of a twisted hyper, or some non-Lagrangian 3D $\cN=4$ SCFT (in the spirit of \cite{Gang:2018huc,Gang:2021hrd}).

\subsubsection{$SU(2)$ with two fundamental hypers}
The third and final theory with the ``$D_2$'' Coulomb branch is an $SU(2)$ theory with two fundamental full hypers, which was discussed in \cite[Section 4.2]{Assel:2018exy} and \cite[Section 7.3.1]{Giacomelli:2024laq}. Here again the bubbling coefficient $c=0$, and the relations between $X=M^2 + M^{-2}$, $Y=\Phi(M^2 - M^{-2})$ and $\Phi^2$ are:
\begin{align}\label{eq:XPh-comm12}
X\Phi^2 - \Phi^2 X &=  -\frac{4i}{r}Y - \frac{4}{r^2} X,\\
\label{eq:YPh-comm12}Y\Phi^2 - \Phi^2 Y &=  -\frac{4i}{r} \Phi^2 X - \frac{4}{r^2} Y,\\
\label{eq:XY-comm12}XY - YX &= -\frac{2i}{r} X^2 + \frac{i}{2r},\\
\label{eq:Y2toY12}Y^2 &= \Phi^2 X^2 - \frac{2i}{r} YX - \frac14 \Phi^2 + \frac{1}{4r^2}.
\end{align}
The commutative limit, $Y^2 = Z X^2 - \frac14 Z$, is, up to rescaling, our familiar complex surface. The full non-commutative algebra $\cA_C$, though, is yet again different.

As for the Higgs branch, the hyper-K\"{a}hler quotient $\C^8///SU(2)$ is, in fact, a disjoint union of two $A_1$ cones. A way to think about it is as follows. Two fundamental hypermultiplets is the same as four fundamental half-hypermultiplets, manifesting a flavor group $SO(4)$. Thus $SU(2)_+ \times SU(2)_-$ acts on the half-hypers, and one finds that the Higgs branch is given by the nilpotent cone in $\mathfrak{su}(2)_+\oplus \mathfrak{su}(2)_-$, which is $A_1 \cup A_1$. Though from the viewpoint of the hyper-K\"{a}hler quotient, it appears like these cones touch at the tip, it is clear from the physical picture that the quantum reality is different: they are attached to the two singular points on the Coulomb branch. Thus, both SCFTs sitting at the singular points have $\cM_H = \cM_X = A_1$ singularity, consistent with the results of \cite{Assel:2018exy} and \cite[Section 7.3.1]{Giacomelli:2024laq}, who used their methods to identify SCFT at the singularity as the IR limit of SQED${}_2$ (i.e., $U(1)$ with two charge-one flavors). This is in contract to our spin-$\frac32$ example, where singularities do no represent intersection with the Higgs branch and support an undetermined SCFT. It would be interesting to develop methods for determining the latter.

\subsubsection{$SU(2)$ with full adjoint hypermultiplet}
An $SU(2)$ gauge theory with a single adjoint full hyper has a nonzero bubbling term with $c^2=\frac1{4r^4}$. To fix the sign of $c$, we may either view this as a $U(2)$ gauge theory (in which the $b=1$ monopole is minuscule), in which the topological $U(1)_{\rm top}$ is gauged, or, even more transparently: Notice that the matter content with only integral weights (such as the adjoint matter) allows for the gauge group $SO(3)$. The $SU(2)$ and $SO(3)$ gauge theories are of course related to each other by gauging the $\mathbb{Z}_2$ symmetry. The $SO(3)$ theory allows for the $b=1$ monopoles (which again are minuscule and do not bubble), so we have $x=M^1 + M^{-1}$, $y=\Phi(M^1 - M^{-1})$ and
\begin{align}
M^1   &= \frac{ir\Phi + \frac12}{ i r \Phi} e^{-\frac{i}{r}\partial_\Phi},\\
M^{-1}&= \frac{ir\Phi-\frac12}{ i r \Phi} e^{\frac{i}{r}\partial_\Phi}.
\end{align}
By computing $x^2 = M^2 + M^{-2} + \frac{-\frac{1}{2r^2}}{\Phi^2 + \frac1r^2} + 2$, we see that indeed this fixes the sign, $c=-\frac{1}{2r^2}$. We first compute relations in the $SO(3)$ theory:
\begin{align}
x\Phi^2 - \Phi^2 x &= -\frac{2i}{r} y - \frac1{r^2} x,\\
y\Phi^2 - \Phi^2 y &= -\frac{2i}{r}\Phi^2 x - \frac1{r^2} y,\\
xy - yx &= -\frac{i}{r} x^2 + \frac{4i}{r},\\
y^2 &= \Phi^2 x^2 -4\Phi^2 -\frac{i}{r} yx + \frac1{r^2}.
\end{align}
In the $SU(2)$ theory, we instead have $X = x^2-2$, $Y=yx$ as the elementary monopoles, and the relations are:
\begin{align}
X\Phi^2 - \Phi^2 X &= -\frac{4i}{r}Y - \frac{4}{r^2} X,\\
Y\Phi^2 - \Phi^2 Y &= -\frac{4i}{r}\Phi^2 X - \frac{4}{r^2} Y -\frac{2i}{r^3},\\
XY - YX &= -\frac{2i}{r} X^2 + \frac{8i}{r},\\
Y^2 &= \Phi^2 X^2 - 4\Phi^2 - \frac{2i}{r}YX + \frac1{r^2} X + \frac{2}{r^2}.
\end{align}
We see that the commutative limits of the $SU(2)$ and $SO(3)$ theories give the same complex surface $y^2 = zx^2 - 4z$ (not even rescalings are required), which is related to the existence of the following automorphism:
\begin{equation}
(x,y,z) \mapsto (x^2-2, yx, z).
\end{equation}
At the same time, the noncommutative algebras are different: We constructed $\cA_C^{SU(2)}$ as a proper inclusion in $\cA_C^{SO(3)}$.

As for the rest of vacua, this theory has no pure Higgs branch. Instead, the Higgs and Coulomb branches are unified into the mixed moduli space, which is typical for enhanced SUSY (indeed, this example has $\cN=8$). What we have found so far (the ``Coulomb branch'') is a slice of the moduli space where the hypermultiplet scalars vanish, $Q=\tQ=0$. If we include $(Q, \tQ)$, we find the total moduli space of the 3D $\cN=8$ $SU(2)$ super Yang-Mills, which is known to be $\C^4/\Z_2$ \cite{Seiberg:1997ax,Gang:2011xp}. If we denote coordinates on $\C^4$ as $(m ,\Phi, Q, \tQ)$, with the identification $(m ,\Phi, Q, \tQ)\sim (-m ,-\Phi, -Q, -\tQ)$, then by choosing $x=e^m + m^{-m}$, $y=\Phi(e^m - e^{-m})$, $z=\Phi^2$, we indeed get $y^2=z x^2 - 4z$ inside the total $\C^4/\Z_2$.

\subsection{Quiver of $SU(2)$ groups}
\begin{figure}[h]
	\centering
	\includegraphics[width=0.4\textwidth]{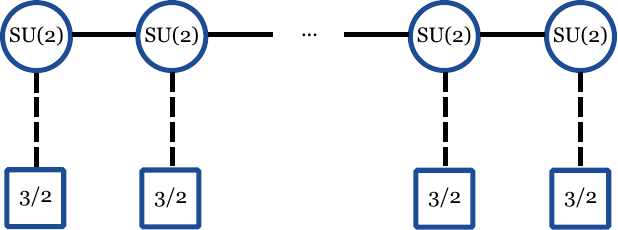}
	\caption{Quiver of $SU(2)$ groups with half-hyper flavors connected by the dashed lines.}\label{fig:quiver}
\end{figure}

Another example we briefly consider is a linear quiver of $SU(2)$ groups, with $n$ gauge nodes, bifundamentals connecting adjacent $SU(2)$ nodes, and one flavor of spin-$\frac32$ half-hyper attached to each node, see Figure \ref{fig:quiver}. Thanks to the presence of this flavor, the bubbling coefficients $c_i$ (which are associated to each node) all vanish.
\\
Thus we have:
\begin{align}
X_i &= M^2_i + M^{-2}_i,\\
Y_i &= \Phi_i(M^2_i - M^{-2}_i),
\end{align}
where the shift operators are:
\begin{align}
M^2 &= \frac1{r^4} \left(\frac12 + \frac{ir}{2}\Phi_i + \frac{ir}{2}\Phi_{i-1}\right)\left(\frac12 + \frac{ir}{2}\Phi_i - \frac{ir}{2}\Phi_{i-1}\right)\left(\frac12 + \frac{ir}{2}\Phi_i + \frac{ir}{2}\Phi_{i+1}\right)\left(\frac12 + \frac{ir}{2}\Phi_i - \frac{ir}{2}\Phi_{i+1}\right) \cr &\times \frac{\left(\frac12 +\frac{ir}{2}\Phi_i\right)\left(\frac12 +\frac{3ir}{2}\Phi_i\right)_3}{(ir\Phi_i+1)ir\Phi_i} e^{-\frac{2i}{r}\partial_\Phi},\\
M^{-2} &= \frac1{r^4} \left(\frac12 - \frac{ir}{2}\Phi_i + \frac{ir}{2}\Phi_{i-1}\right)\left(\frac12 - \frac{ir}{2}\Phi_i - \frac{ir}{2}\Phi_{i-1}\right)\left(\frac12 - \frac{ir}{2}\Phi_i + \frac{ir}{2}\Phi_{i+1}\right)\left(\frac12 - \frac{ir}{2}\Phi_i - \frac{ir}{2}\Phi_{i+1}\right) \cr &\times \frac{1}{(ir\Phi_i-1)ir\Phi_i} e^{+\frac{2i}{r}\partial_\Phi}.
\end{align}
One can then derive all the relations precisely in the same way. We only derive the commutative limit of the Coulomb branch, written in terms of $X_i$, $Y_i$ and $Z_i=\Phi_i^2$:
\begin{equation}
Y_i^2 = Z_i X_i^2 - \frac{27}{1024} Z_i(Z_i-Z_{i-1})^2 (Z_i-Z_{i+1})^2, \quad i=1\dots n.
\end{equation}
Here we set $Z_0=Z_{n+1}=0$. This system of equation is an answer to the Coulomb branch, but we could of course extract much more (the quantum ring $\cA_C$ and the correlators, since this is a good theory).

\section{Outlook}
In this work, we completed the program of \cite{Dedushenko:2016jxl,Dedushenko:2017avn,Dedushenko:2018icp} by constructing the quantized Coulomb branch $\cA_C$ of a general Lagrangian 3D $\cN=4$ gauge theory that has half-hypermultiplets (i.e., a \emph{noncotangent} theory). When the theory is good, we also have the trace map on the algebra:
\begin{equation}
(\cO \in \cA_C) \mapsto \langle Z_{HS^3}| \cO |Z_{HS^3}\rangle = \frac1{|\cW|} \sum_{B\in\Lambda^\vee} \int_{\mathfrak{t}} \dd^r\sigma\, \mu(\sigma, B) Z_{HS^3}(\sigma, B, \mathbb{L}) \cO Z_{HS^3}(\sigma, B, \mathbb{L}),
\end{equation}
where $Z_{HS^3}$ is the empty hemisphere partition function \eqref{eq:emptyZHS3}, $\mu(\sigma, B)$ is the gluing measure defined in \eqref{eq:gl-measure}, and $\cO$ is the shift operator representing an element of $\cA_C$. This map allows to compute correlation functions in the topological quantum mechanics associated to the Coulomb sector, which requires nontrivial integration over $\sigma$. However, we get the algebra $\cA_C$ itself (the quantized Coulomb branch) essentially for free, without any integration.

In the answer provided in the main body of the paper, we ignored masses and FI parameters, however, they can be easily incorporated. To include masses, one simply replaces $w\cdot\Phi$ by $w\cdot\Phi + M$ inside the Pochhammer symbols in the numerators of shift operators. This will have the effect of deforming the Coulomb branch by complex masses. The FI terms are a bit trickier since they lift the Coulomb branch. As explained in \cite[Section 5.1.2]{Dedushenko:2017avn}, the FI terms introduce explicit $\varphi$-dependence $e^{r \zeta b \varphi}$ into the monopole operator, and they also multiply the gluing measure by $e^{-8\pi^2ir\zeta\cdot\sigma}$.

There are still some problems left to solve in this area. A few notable ones:
\begin{enumerate}
	\item Proving the polynomiality conjecture in general.
	\item Developing more efficient techniques for extracting the answers from our formalism, in particular, the bubbling coefficients and the generators and relations of $\cA_C$.
	\item Developing tools for extracting the data of SCFT at the singularity of moduli space. Our example of $SU(2)$ with one spin-$\frac32$ half-hyper is an interesting test case. This theory is quite rigid: it has no mass deformations in the UV. Its Coulomb branch has two $A_1$ singularities supporting some SCFT (which has $A_1$ Coulomb branch and no Higgs branch). At present, we do not know how to extract this SCFT, and the tools of \cite{Assel:2017jgo,Assel:2018exy,Giacomelli:2023zkk,Giacomelli:2024laq} do not seem to apply. This is analogous to how Argyres-Douglas theories appear in 4D at the special singular loci of the Coulomb branch.
\end{enumerate}

\section*{Acknowledgements}
MD thanks Silviu Pufu and Matteo Sacchi for correspondence. DR would like to thank Martin Rocek for correspondence and discussion during the writing of this paper.
\appendix

\section{Correlators}\label{app:corr}
The goal of this appendix is to show that the monopole-antimonopole correlation function is independent of the choice of Lagrangian splitting. We wish to compute
\begin{equation}
\langle \cM^2\cM^{-2} \rangle  = \frac{1}{|\mathcal W|Z_{S^3}} \sum_{b\,\in\, \Lambda^{\vee}} \int_{\mathfrak t} \dd\sigma\,\mu(\sigma, B, \mathbb L) Z_{HS^3} \cM^2\cM^{-2}Z_{HS^3}(\sigma, B, \mathbb L).
\end{equation}
We will restrict, as with the rest of these examples, to $SU(2)$, but for now we are leaving the choice of matter content of the theory general. Any choice of Lagrangian splitting for a given set of half-hypermultiplets will 'split' the weights into $\mathbb L$ and $\mathbb L^{\perp}$. For a given theory, we can define $\mathbb L$ as
\begin{equation}
\mathbb L = \left\{w_1^{I_1}, ... , w_k^{I_k}, -w_1^{I'_1}, ..., -w_l^{I'_l}\right\}
\end{equation}
with weights $w_i$ of multiplicity $I_i$ and negative weights $-w_i$ with multiplicity $I'_i$. The weights are indexed by $k, l$ as they can, in principle, be different numbers, but in many cases (increasing with the complexity of the theory), these numbers are equal. These multiplicities are constrained by the relations
\begin{equation}
|I_i - I_i'| \leq \text{max}(I_i, I_i'), \quad I_i + I_i' = \mathcal I_i
\end{equation}
where $\mathcal I_i$ is the multiplicity of a given $|w_i|$, i.e, the multiplicity if all of a given $w_i$ were positive (or negative). The repeated index refers to selecting a weight $w_i$ and its reflection $-w_i$. With this notation, we are then free to partition the multiset $\mathbb L$ in the following way:
\begin{equation}
\mathbb L^+ = \left\{w_1^{I_1}, ... , w_k^{I_k}\right\}, \quad \mathbb L^- = \left\{-w_1^{I'_1}, ..., -w_l^{I'_l}\right\}.
\end{equation}
This groups all positive weights of $\mathbb L$ together as well as their reflections in two distinct sets. Together with the constraints on $I_k$ and $I'_l$, this guarantees that the intersection $\mathbb L^+\,\cap\,\mathbb L^-$ is empty. This is useful, primarily, because we may re-express the sum of weights in terms of these two sets:
\begin{equation}\label{weightsum}
A \equiv \sum_{w\in\mathbb L} |w| = \sum_{w\in\mathbb L^+} I_iw_i - \sum_{w'\in \mathbb L^-} I_j'w'_j
\end{equation}
with the minus sign included to account for the minus sign in the definition of the elements of $\mathbb L^-$. We can further re-express the shift operators in this new language
\begin{equation}
M^2 = \frac{(-1)^{\sum_{w_k} I_kw_k}}{r^{A-2}(ir\Phi + 1)(ir\Phi)} \prod_{w_k\in \mathbb L^+}\prod_{j = 0}^{2w_k -1}\left(\frac 12 + irw_k\Phi + j\right)^{I_k}e^{-\frac{2i}{r}\del_{\Phi}}
\end{equation}
\begin{equation}
M^{-2} = \frac{(-1)^{-\sum_{w'_k} I'_kw'_k}}{r^{A-2}(ir\Phi - 1)(ir\Phi)} \prod_{w'_k\in \mathbb L^-}\prod_{j = 0}^{-2w'_k -1}\left(\frac 12 - irw'_k\Phi + j\right)^{I'_k}e^{\frac{2i}{r}\del_{\Phi}}.
\end{equation}
In these expressions, the pre-factor contains the contribution from the sign function written in terms of $\mathbb L^+$ and $\mathbb L^-$, respectively\footnote{The set symbols have been left out of the sign functions for notational convenience.}, as well as the explicitly evaluated vector-multiplet denominators. The product then selects a particular weight in $\mathbb L^+$ and its multiplicity and then expands the Pochhammer symbol $I_k$ times. We observe that such an expression of the shift operators is possible on account of the observation that in \ref{eq:shift-op}, only combinations of $wb$ appear together--- either depending on the absolute value of $wb$ (which corresponds to A in the expression, as it is agnostic of the sign of the weights) or only contributing if $wb$ is positive. Passing to the antimonopole from the monopole shift operator flips the accessed weights in $\mathbb L$.

Now we may begin to compute the monopole-antimonopole two point function. To reorient, our ultimate goal is to show that this two point function does not depend on the choice of $\mathbb L$. The inspiration for the re-expression of the following above shift operators is that, in their expression, not every weight from $\mathbb L$ contributes. This is the sign of Lagrangian splitting--- when the choice of sign in $\mathbb L$ "skips over" negatively charged weights in the monopole shift operator and instead places them in the antimonopole shift operator. In other words, the individual shift operators \textit{do} depend on Lagrangian splitting. We observe, however, that the \textit{product} of two oppositely charged shift operators (like what enters in the two-point function) need not depend on Lagrangian splitting.

For the hemisphere partition function and gluing measure, no redefinition is necessary, and this can be seen from the definition of \ref{eq:abelian-monopole-contribution} and \ref{eq:gl-measure}, respectively. From their form, it can be seen that it does not care which elements of $\mathbb L$ it runs over--- for them, the choice of Lagrangian splitting is truly arbitrary. Thus, we need only show that the dependence on $\mathbb L^+$ and $\mathbb L^-$ drop out of the correlator.

The following terms survive flux conservation:
\begin{equation}\label{weylaveragedfunction}
\langle \cM^2\cM^{-2}\rangle = \langle M^2M^{-2}\rangle + \langle M^{-2}M^2\rangle.
\end{equation}
Focusing on the first term on the right-hand side of \ref{weylaveragedfunction}, we first act $M^{-2}$ on the hemisphere partition function $HS^3$, followed by $M^2$, which yields a correlator of the following form
\begin{equation}\label{abeliancorrelator}
\langle M^2M^{-2}\rangle = \frac{\pi^{2-\mathcal I}}{2Z_{S^3}}\int_{\mathfrak t}\dd\sigma\,\mathcal{V}(\sigma)\mathcal R(\sigma)\frac{\sinh^2(\pi\sigma)}{\prod_{w\in\mathbb L}\cosh(\pi |w|\sigma)}.
\end{equation}
Here, $\mathcal I$ is the size of $\mathbb L$, namely
\begin{equation}
\mathcal I = \sum_{i = 1}^k I_i + \sum_{j = 1}^l I'_j
\end{equation}
for each positive and negative weight. It also depends on two functions $\mathcal{V}(\sigma)$ and $\mathcal{R}(\sigma)$, whose forms are
\begin{equation}\label{vfunctioncor}
\mathcal{V}(\sigma) = \frac \sigma r\frac{(-1)^{\sum_{w_k} I_kw_k -\sum_{w'_k} I'_kw'_k}}{r^{A-4}}\frac{1}{i\sigma(1+i\sigma)^2(i\sigma + 2)}
\end{equation}
and
\begin{equation}\label{rfunctioncor}
\mathcal{R}(\sigma) = \bigg(\prod_{w_k\in \mathbb L^+}\prod_{j = 0}^{2w_k -1}\left(\frac 12 + irw_k\Phi + j\right)^{I_k}\bigg)\bigg(\prod_{w'_k\in \mathbb L^-}\prod_{j = 0}^{-2w'_k -1}\left(\frac 12 - irw'_k\Phi + j\right)^{I'_k}\bigg).
\end{equation}
Let us address \ref{abeliancorrelator} first. The rightmost factor in the two-point function is the ultimate fate of the gluing measure and the hemisphere partition functions--- with the numerator coming from the vector-multiplet contribution, and the denominator the half-hypermultiplet content. It depends on the absolute value of a given weight, and therefore obviously does not depend on Lagrangian splitting. Turning to \ref{vfunctioncor}, We see that the only dependence on $\mathbb L^+$ and $\mathbb L^-$ is in the sign function. But due to the relation \ref{weightsum}, this too is independent of Lagrangian splitting. We conclude that $\mathcal{V}(\sigma)$ does not depend on Lagrangian splitting. Finally, we look at \ref{rfunctioncor}. We need to show that this generates every possible term from $\mathbb L$. To see this, we note that because $\mathbb L^+$ and $\mathbb L^-$ partition $\mathbb L$, their union forms all of $\mathbb L$. One can observe that $\mathcal R(\sigma)$ runs over every element of $\mathbb L^+$ and $\mathbb L^-$, and thus doesn't care about the Lagrangian splitting. We conclude that the correlator is then independent of the choice of Lagrangian splitting.

\bibliographystyle{utphys}
\bibliography{bibliography}
\end{document}